\documentclass{nature}

\usepackage{bm}
\usepackage{graphicx,epsfig}
\usepackage{amsmath,amssymb,mathtools}
\usepackage[rgb]{xcolor}
\usepackage{hyperref}
\def\be{\begin{equation}}
\def\ee{\end{equation}}

\def\bc{\begin{center}}
\def\ec{\end{center}}
\def\bea{\begin{eqnarray}}
\def\eea{\end{eqnarray}}
\newcommand{\avg}[1]{\langle{#1}\rangle}

\def\ie{\textit{i.e.}}

\makeatletter
\let\saved@includegraphics\includegraphics
\AtBeginDocument{\let\includegraphics\saved@includegraphics}
\renewenvironment*{figure}{\@float{figure}}{\end@float}

\makeatother

\title{Triadic percolation induces dynamical topological patterns in higher-order networks }

\author{Ana P. Mill\'an$^{1}$, Hanlin Sun$^{2}$, Joaqu\'in J. Torres$^{1}$ \& Ginestra Bianconi$^{3,4}$}

\begin{document}
\maketitle

\begin{affiliations}
\item Institute ``Carlos I'' for Theoretical and Computational Physics, and Electromagnetism and Matter Physics Department, University of Granada, E-18071 Granada, Spain
\item Nordita, KTH Royal Institute of Technology and Stockholm University, Hannes Alfvéns väg 12, SE-106 91 Stockholm, Sweden
\item Centre for Complex Systems, School of Mathematical Sciences, Queen Mary University of London, London E1 4NS, UK
\item The  Alan  Turing  Institute,  96  Euston  Road,  London,  NW1  2DB,  UK
\end{affiliations}

\newpage
\section*{Abstract}

\begin{abstract}
Triadic interactions are higher-order interactions which occur when a set of nodes affects the interaction between   two other nodes.
Examples of triadic interactions are present in the brain when glia modulate the synaptic signals among {neuron pairs or} when interneuron axo-axonic synapses enable presynaptic inhibition and facilitation,
{and} in ecosystems when one or more species can affect the interaction among two other species. On random graphs, triadic percolation has been recently shown to turn percolation into a fully-fledged dynamical process in which the size of the giant component undergoes a route to chaos. However, in many real cases, triadic interactions are local and occur on spatially embedded networks. Here we show that triadic interactions in spatial networks induce a very complex spatio-temporal modulation of the giant component which gives rise to triadic percolation patterns with  significantly different topology. We classify the observed patterns (stripes, octopus and small clusters) with topological data analysis and we assess their information content (entropy and complexity). Moreover we illustrate the multistability of the dynamics of the triadic percolation patterns and we  provide a comprehensive phase diagram of the model.
These results open new perspectives in percolation as they demonstrate that in presence of spatial triadic interactions, the giant component can acquire a time-varying topology. Hence, this work provides a theoretical framework that can be applied to model realistic scenarios in which the giant component is time-dependent as in neuroscience. 
\end{abstract}

\newpage

\section{Introduction}
Percolation \cite{dorogovtsev2008critical,cohen2010complex,li2021percolation,lee2018recent,d2019explosive,gao2015recent,shekhtman2016recent} is a pivotal critical phenomenon that captures the non-linear response of a network to the damage of its links. 
Thus, percolation has a wide range of applications in complex systems including brain and infrastructure networks \cite{reis2014avoiding,morone2015influence,danziger2016effect,bonamassa2019critical}. The dynamical model usually implied by the percolation processes describes cascades or avalanches of failure events,  typically reaching a static absorbing state, which might result in a  dismantled network.
Hence the traditional approach to percolation fails to describe the situation encountered for instance in neuroscience where their giant component  of functional brain networks dynamically varies in time without reaching a static configuration.
In order to capture  this scenario, higher-order triadic interactions  have recently been shown to be a key, leading to the formulation of  triadic percolation \cite{sun2023dynamic}.

Higher-order networks \cite{bianconi2021higher,battiston2020networks,battiston2021physics,torres2021and, bick2023higher,majhi2022dynamics} include interactions between two or more nodes. These generalized network structures are attracting large interest because they are transforming significantly our understanding of the interplay between topology and dynamics of complex networks.
Indeed the structure of higher-order networks captures the underlying topology and geometry of the data, with applications from ecology \cite{grilli2017higher,silk2022capturing, bendick2020topological} to cancer research \cite{nicolau2011topology, stolz2021topological, stolz2022multiscale}. 
Evidence of higher-order coupling appears in several natural systems \cite{bianconi2021higher,battiston2020networks,battiston2021physics,torres2021and, bick2023higher,majhi2022dynamics}, including brain dynamics, chemical interaction networks, and the climate \cite{giusti2016two, faskowitz2022edges, jost2019hypergraph, su2022climatic,boers2019complex, rosenthal2018mapping, varley2023partial, clauw2022higher}. 
Higher-order interactions have the potential to dramatically change the emergent behavior of a dynamical model \cite{bianconi2021higher,majhi2022dynamics}, as evidenced in percolation \cite{sun2023dynamic,santos2019topological,bobrowski2020homological,lee2021homological}, synchronization \cite{millan2020explosive, skardal2019abrupt, zhang2021unified, mulas2020coupled}, diffusion \cite{torres2020simplicial,carletti2020random},  game theory \cite{alvarez2021evolutionary} or contagion dynamics \cite{st2021universal, de2020social,iacopini2019simplicial}.

Triadic interactions \cite{grilli2017higher,sun2023dynamic,bairey2016high,wang2009genome} are higher-order interactions that occur when one or more nodes regulate the interaction between  two other nodes. This is the role for instance of glia cells in neuronal networks, as they modulate synaptic interactions between neuron pairs \cite{cho2016optogenetic}. In the brain,  presynaptic inhibition
and facilitation is a further example of triadic interactions.
Indeed, axo-axonic synapses involve interneurons that induce presynaptic inhibition or facilitation, which reduce or enhance, respectively, postsynaptic response \cite{kandelbook,Byrne425}. 
Similarly, in ecological systems, the presence of a third species may modulate the interaction between two given species \cite{grilli2017higher,bairey2016high}. 

Recently, in Ref. \cite{sun2023dynamic}, it was shown that triadic percolation, in which triadic interactions up- or down-regulate links, defines a fully-fledged dynamical process in which the giant component becomes dynamical,  and its size undergoes a period-doubling and a route to chaos. 
However, Ref. \cite{sun2023dynamic} only addresses triadic percolation on a random graph while in a large variety of cases, real-world networks have a geometric embedding, such is the case for instance for neuronal networks \cite{fornito2016fundamentals, breakspear2017dynamic}, communication and transportation systems \cite{lambiotte2008geographical}.
For many of these systems, an exponential wiring cost fits the observed spatial distribution of link lengths \cite{ercsey2013predictive,guamuanuct2018mouse,ganti2009spatial,halu2014emergence,danziger2016effect,bullmore2012economy,danziger2022recovery,markov2014weighted,horvat2016spatial,deco2021rare,watts2002identity}.
Despite many advances in the study of percolation on spatial random networks and lattices \cite{shekhtman2016recent,berezin2015localized,fan2020universal,dong2020measuring,danziger2016effect,bonamassa2019critical,gross2017multi,gao2015recent}, their analysis in the presence of a geometric embedding remains an important theoretical challenge. 

In this work we address the study of triadic percolation defined on spatially embedded networks with triadic interactions. 
We show that, since in these networks interactions are local, triadic percolation induces a spatial structure on the giant component, leading to triadic percolation patterns, which display a distinct time-varying geometry and topology. 
Here we investigate the spatio-temporal complexity of the observed dynamics of the triadic percolation patterns, which goes beyond the basic statistics and dynamics of the size of the giant component. 
To explore these findings, we investigate the spatial properties of the triadic percolation patterns with Topological Data Analysis (TDA) \cite{ghrist2008barcodes,otter2017roadmap,vaccarino2022persistent,curto2017can}, and information theory tools \cite{sigaki2018history,bandt2002permutation,ribeiro2012complexity}, identifying the emergence of three distinct types of patterns: small clusters, stripes and octopus (or complex) patterns. 
We launch an in-depth numerical investigation of the dynamics of these patterns: we show evidence of intermittency with patterns of different types occurring in the same time-series, and we provide evidence of periodic (blinking) spatio-temporal behavior of the patterns. These analyses allow us to reveal the topological and dynamical nature of the phase diagram of triadic percolation in spatial networks.

Triadic percolation on spatial networks captures basic mechanisms that play in a large variety of spatially embedded systems with triadic interactions, notably including brain networks. In order to reveal its implications, we kept our model very general. At the same time, the stylized nature of this model proposes a  framework that combines percolation theory, topology and non-linear dynamics that can become the starting point for more realistic models of specific scenarios in which the giant component is strongly time-dependent.

\begin{figure}[!tbh]
    \centering \linespread{1.1} \selectfont
    \linespread{1.1}
    \selectfont
\includegraphics[width=0.6\columnwidth]{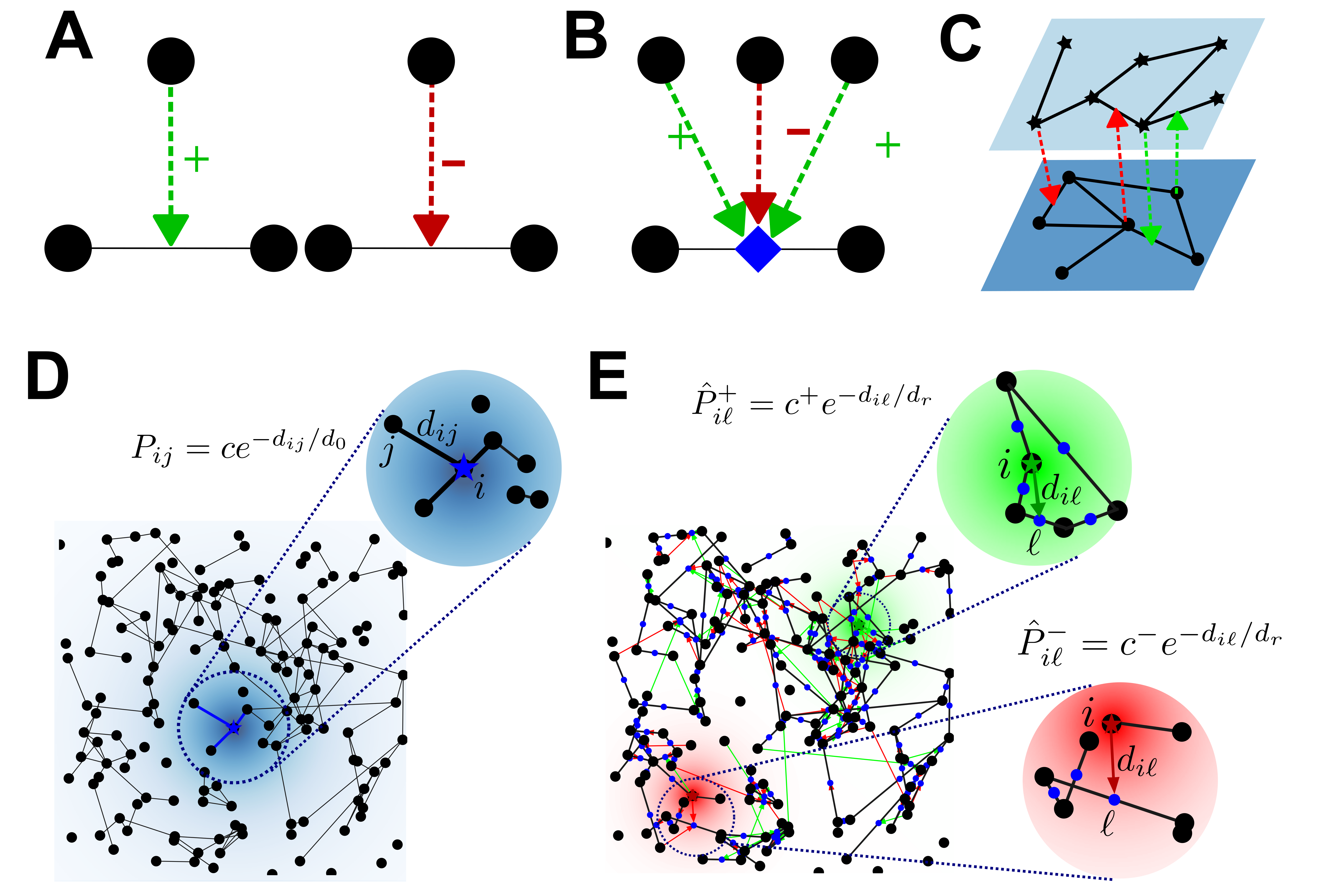}     
    \caption{Network and triadic percolation models.
    \textbf{A} Illustration of the triadic interaction in which one node regulates the link between two other nodes with a positive (left image) or negative (right image) effect. 
    \textbf{B} Triadic interactions allow for several regulator nodes for each link. In this case the presence of a link is determined by a function taking into account both positive and negative regulatory interactions.
    \textbf{C} Triadic two-layer network, where each layer regulates the interactions in the other layer. 
    \textbf{D} Illustration of the spatial structural network. We highlight the local connectivity mechanism in the zoomed area. Two nodes $i$ and $j$ are connected by a structural link with a probability $P_{ij}$ that depends on the Euclidean distance $d_{ij}$ between the two nodes, on the typical length for structural links $d_0$, and on the parameter $c$ that determines the average degree of the structural network. 
    The probability is schematically indicated by the shaded blue color in the top zoomed area. 
    \textbf{E} Illustration of the spatial network with triadic interactions. The higher-order network is formed by the spatial structural network shown in panel (D) and the triadic interactions. The higher-order network with triadic interactions is formed by nodes (black circles), structural links (black lines 
    the center of each structural link, indicating its location, is shown by smaller blue circles), and triadic regulatory interactions which can either be positive (green arrows) or negative (red arrows) as depicted in the zoomed areas. A positive regulatory interaction between a node $i$ and a structural link $\ell$ is added with a probability $\hat{P}^+_{i \ell}$,  schematically indicated by the shaded green color, while a negative regulatory interaction is added with probability $\hat{P}^-_{i \ell}$, schematically indicated by a shaded red color.  The probabilities $\hat{P}^{+}_{i \ell}$ and $\hat{P}^{-}_{i \ell}$ depend on the Euclidean distance $d_{il}$ between a node and the link it regulates, on the typical range $d_r$ for regulatory links, and on the parameters $c^{+}$ and $c^{-}$ determining the average degree of positive and negative regulatory interactions respectively.
    }
\label{fig:figure1_spatialillustration}
\end{figure}

\section{Triadic percolation on spatial networks with triadic interactions}
Triadic interactions are higher-order interactions that occur when one node regulates the interaction between  two other nodes, either positively or negatively (see Fig. $\ref{fig:figure1_spatialillustration}$A,B). 
Positive regulations facilitate the interaction between the two nodes while negative regulations inhibit their interaction. 
Interestingly, triadic interactions can also exist in a multilayer setting \cite{bianconi2018multilayer} when the regulator nodes are of a different type (see Fig. $\ref{fig:figure1_spatialillustration}$C). A typical example of this latter type of triadic interaction is present in neuronal systems, where glia cells, for example, can either facilitate or inhibit the synaptic signals between neurons. Another example is the existence of interneurons that establish axo-axonic synapses that can induce presynaptic facilitation and inhibition of other synapses. 
For the sake of simplicity, in this work we consider the simple scenario in which the regulator nodes belong to the same structural network whose links are regulated by triadic interactions, as it occurs in presynaptic inhibition and facilitation. However the multilayer scenarios can be a natural extension to the proposed framework.

We consider a higher-order spatial network with triadic interactions embedded in a $2D$ torus (square of size $L$ with periodic boundary conditions). This higher-order network can be thought of as a multilayer network \cite{bianconi2018multilayer}  formed by  a structural network $G_s$ and a regulatory network $G_r$. 
The structural network $G_s=(V, E)$ comprises the set $V$ of nodes and the set $E$ of structural links between them. We indicate with $N$ the number of nodes in $V$, i.e. $N=|V|$.
The regulatory network $G_r=(V, E, W),$ on the other hand, is a  signed factor network (bipartite network) formed by the nodes in $V$ (called regulators), the structural links in  $E$ (playing the role of factor nodes), and the signed regulatory interactions between them, specified in the set $W$. 
Note that the signs of regulatory interactions are considered here as a property of the regulatory link rather than the regulator node. 
This means that a given regulator node can positively regulate some links (thus called positive regulator to these links) and negatively regulate some other links (thus called negative regulator to these links).

The spatial structural network is constructed according to the Waxman model \cite{waxman1988routing} (see Fig. $\ref{fig:figure1_spatialillustration}$). This model accounts for local interactions as links between pairs of nodes are drawn  with a probability that decays exponentially with their distance as found in several brain structural investigations \cite{ercsey2013predictive,guamuanuct2018mouse}.
Similarly, our model of spatial triadic interactions also establishes regulatory interactions among pairs of nodes and structural links with a probability that decays exponentially with their distance (See Methods for details of the model).
The model depends on few parameters: $d_0$ and $d_r$ indicate the typical range of interactions for structural and regulatory links respectively, while the positive parameters $c$, $c^+$ and $c^-$  can be further used to modulate the average structural ($c$) or regulatory (positive-$c^+$- and negative-$c^-$-) degree of the structural links (number of regulator nodes).

The triadic interactions can activate or deactivate the structural links giving rise to triadic percolation. 
In triadic percolation, the activity of structural links that controls the network connectivity is determined by the triadic interactions that regulate them. Conversely the activity of the nodes is dictated by the network connectivity via the percolation process.

Triadic percolation is defined as follows.
At $t=0$, all structural links are active. 
For $t \geq 1$, the dynamics is given by a simple $2$-step iterative algorithm:
\begin{itemize}
    \item Step $1$: Given the configuration of activity of the structural links at time $t-1$, nodes are considered active if they belong to the largest connected component of the structural network. Otherwise the nodes are considered inactive.
    \item Step $2$: Given the set of all active nodes obtained in Step $1$,  all the links that are connected at least to one active negative regulator node and/or that are not connected to any active positive regulator node are deactivated. The remaining links remain intact only with probability $p$.
\end{itemize}
At each time step $t$, the state of the structural network is given by the binary vector $\bm{s}=\bm{s}(t)$, of elements $s_i(t)$ indicating whether node $i$ is active ($s_i(t)=1$) or inactive ($s_i(t)=0$) at time $t$. The size of the giant component $R$ indicates the fraction of nodes in the giant component,  which in triadic percolation is time dependent.
Note that in triadic percolation  the dynamic is deterministic for $p=1$, while for  $0<p<1$ the dynamic is stochastic, 
\ie, the triadic interactions do not fully determine the dynamic process. 
Thus $p$ acts as the control parameter of the dynamics.

It follows that  spatial triadic percolation includes both quenched and annealed disorder. The quenched disorder is determined by the structure of the higher-order network with triadic interactions, while $p$ drives the annealed disorder and dictates the randomness of deactivation events.

Triadic percolation in spatial networks generates topological patterns of the giant component  with non trivial dynamical evolution. These patterns and their dynamics depend significantly on the control parameter $p$ that determines the highly non trivial phase diagram of triadic percolation in spatial networks, as we discuss below.

\section{Emergence of triadic percolation patterns }
Typically, in percolation theory,  the  percolation transition is described by monitoring  the  size $R$ of the giant component  as a function of the control parameter $p$. 
In spatial triadic percolation we see that this measure fails to capture the complex topological patterns, called triadic percolation patterns,  acquired by the giant component as long as the range of the structural and regulatory interactions is local. 
Hence we need to develop and use additional quantitative methods to classify these patterns and encode their information content.

Visual inspection  of the spatial distribution of the giant component induced by spatial triadic percolation suggests the emergence of three qualitatively different types of patterns as shown in Fig. $\ref{fig:figure2_TDAmethod}$:
\begin{enumerate}
    \item Small clusters 
    of active nodes, of relatively small size. 
    \item Octopus, tubular-like shapes predominantly formed by wide lanes that cross the borders (potentially more than once) and may also form inner loops. 
    \item Stripes, either horizontal or vertical, that reach the borders forming closed loops (due to the periodic boundary conditions) and are of similar width than the octopus. Multiple stripes may also emerge.   
\end{enumerate}
The emergent patterns share a strong spatial structure, with separated active (high density of active nodes) and inactive (null density) areas, leading to a heterogeneous distribution of activity. 
The spatial property is a direct consequence of the local triadic interactions, and it is lost when this requirement is relaxed.

\subsection{TDA classification of triadic percolation patterns}

\begin{figure}[!tbh]
    \centering \linespread{1.1} \selectfont
    \linespread{1.1}
    \selectfont
    \includegraphics[width=0.8\columnwidth]{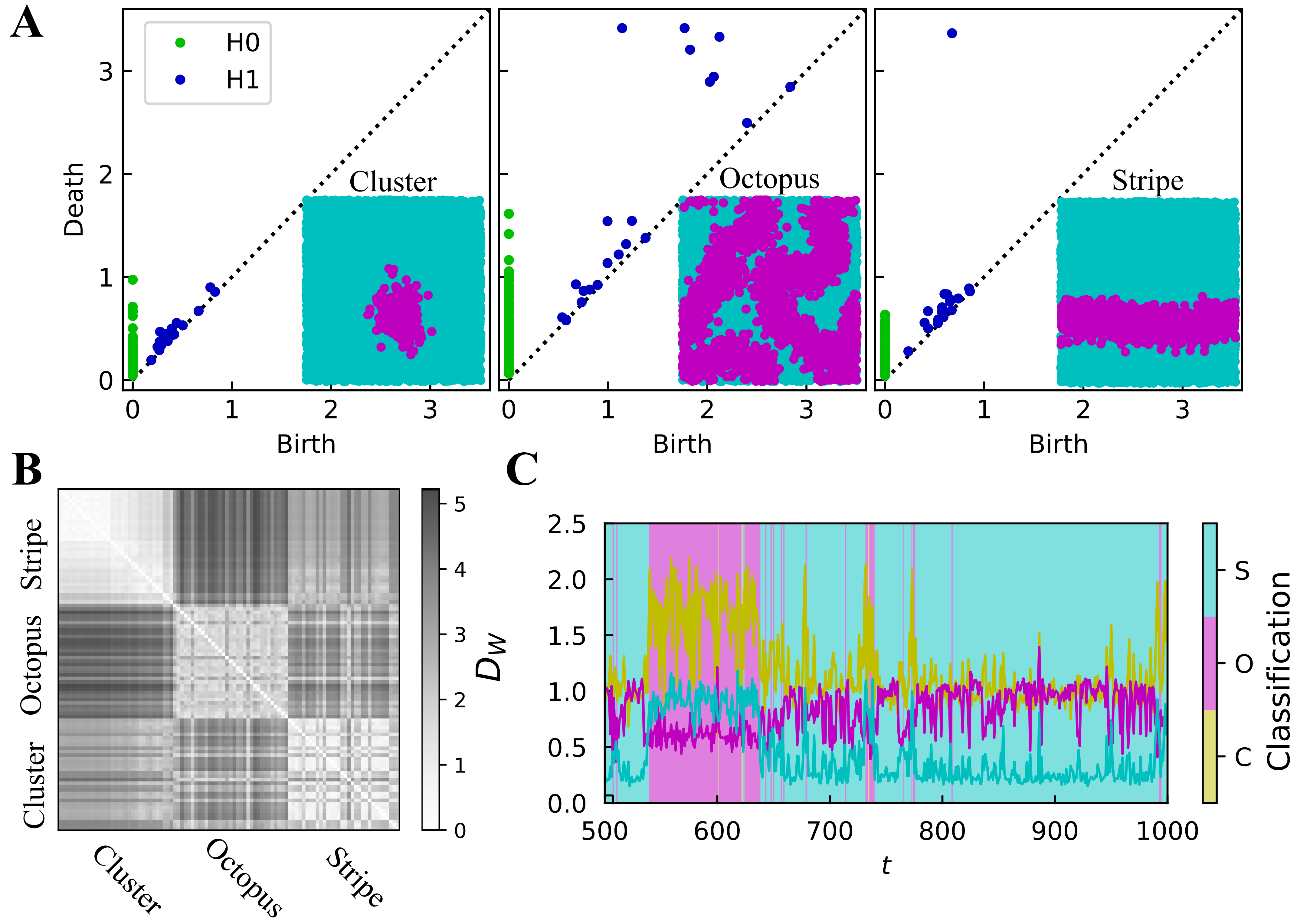}
    \caption{Topological classification of the triadic percolation patterns. 
    \textbf{A} The $0$-homology ($H_0$) and $1$-homology ($H_1$) persistence diagrams  corresponding to an exemplary \emph{Cluster} (C), \emph{Octopus} (O) and \emph{Stripe} (S) patterns, as shown by the insets, where turquoise (pink) dots stand for inactive (active) nodes. 
    Major differences in the persistence diagrams of $H_0$ and $H_1$  can be observed. 
    \textbf{B} Pattern dis-similarity is measured with the Wasserstein distance $D_W$, here  shown  between template patterns corresponding to the Cluster, Octopus and Stripe patterns as indicated by the labels. To perform the classification, $33$ templates of each pattern class have been considered.  Octopus patterns show larger within class topological variability.  
    \textbf{C} Illustrative pattern classification 
    for an exemplary time-series. 
    First, the distance of each state ${\bm s}(t)$ to each pattern class $P$, $P\in \left\lbrace C,O,S \right\rbrace$ is taken to be the the minimum $D_W$ distance between $s(t)$ and the template states of each pattern class (data lines). Then, the triadic percolation pattern at time step $t$ is assigned the closest template pattern (as indicated by the shaded background areas). Here the triadic percolation results are obtained for  $p=0.8$, $N=10^4$, $c^+=c^-=0.2$, $d_0=d_r=0.25$, $c=0.4$, $\rho=100$.}
    \label{fig:figure2_TDAmethod}
\end{figure}

To classify the triadic percolation patterns, we have considered their distinct topology.
Even if clusters and stripes can be of similar or comparable size, the critical difference between them is that stripes reach the borders of the torus and close a loop: whereas clusters have no holes, stripes have one (per stripe in the pattern). 
Conversely, octopus patterns may present more than one macroscopic component (typically of different sizes) and at least one hole. 

TDA, and specifically persistent homology \cite{zomorodian2012topological, centeno2022hands}, is the ideal method to detect these differences.
Persistent homology  encodes a point cloud (in this case the triadic percolation patterns) into a simplicial complex and then tracks the $k$-homological classes. The $k$-homological classes define the topology of the simplicial complex and are  in one-to-one correspondence with the independent $k-$dimensional holes: in dimension $0$, these are connected components; in dimension $1$, cycles (or loops), i.e. one dimensional holes; in dimension $2$, two dimensional holes (as in a triangulation of a sphere), and so on.  Here we consider Vietoris-Rips complexes, built from the point cloud by connecting all points at a spatial (Euclidean) distance smaller than a certain value (radius) and by filling all the cliques of the resulting network. By increasing the radius (i.e. the maximum distance between nodes to be considered connected), the persistent homology diagram indicates the emergence and death of each homological class of the data (see Fig. $\ref{fig:figure2_TDAmethod}$A). The (dis-)similarity between two patterns can then be measured by means of the Wasserstein distance $D_W$ between their respective persistence diagrams (see Methods section for details).

The distance $D_W$ between same-class patterns is much smaller than the between-class distance, as observed in Fig. $\ref{fig:figure2_TDAmethod}$B for hand-selected \emph{template} patterns ($n=33$ of each class). This result quantifies the finding that emergent pattern classes present distinct homology. 
Notably, the Wasserstein distance also captures the larger variability within the octopus class, with larger $D_W$ values. 
Using the set of template patterns, we classified each emergent pattern into one of the three nominal classes by identifying the closest (smallest $D_W$) class to the pattern, 
as illustrated in Fig. $\ref{fig:figure2_TDAmethod}$. 

In this manner, TDA  can provide a quantitative classification of the topological patterns acquired by the giant component in spatial triadic percolation. Hence the spatio-temporal dynamics of the giant component can be considered as a time-series of distinct topological patterns. 
Note however that, as we will reveal in the following, triadic percolation gives rise to a dynamics of triadic percolation patterns that is also largely affected by their geometry (for instance the position of the barycenter of the patterns).  
\begin{figure}[!tbh]
    \centering \linespread{1.1} \selectfont
    \includegraphics[width=0.9\columnwidth]{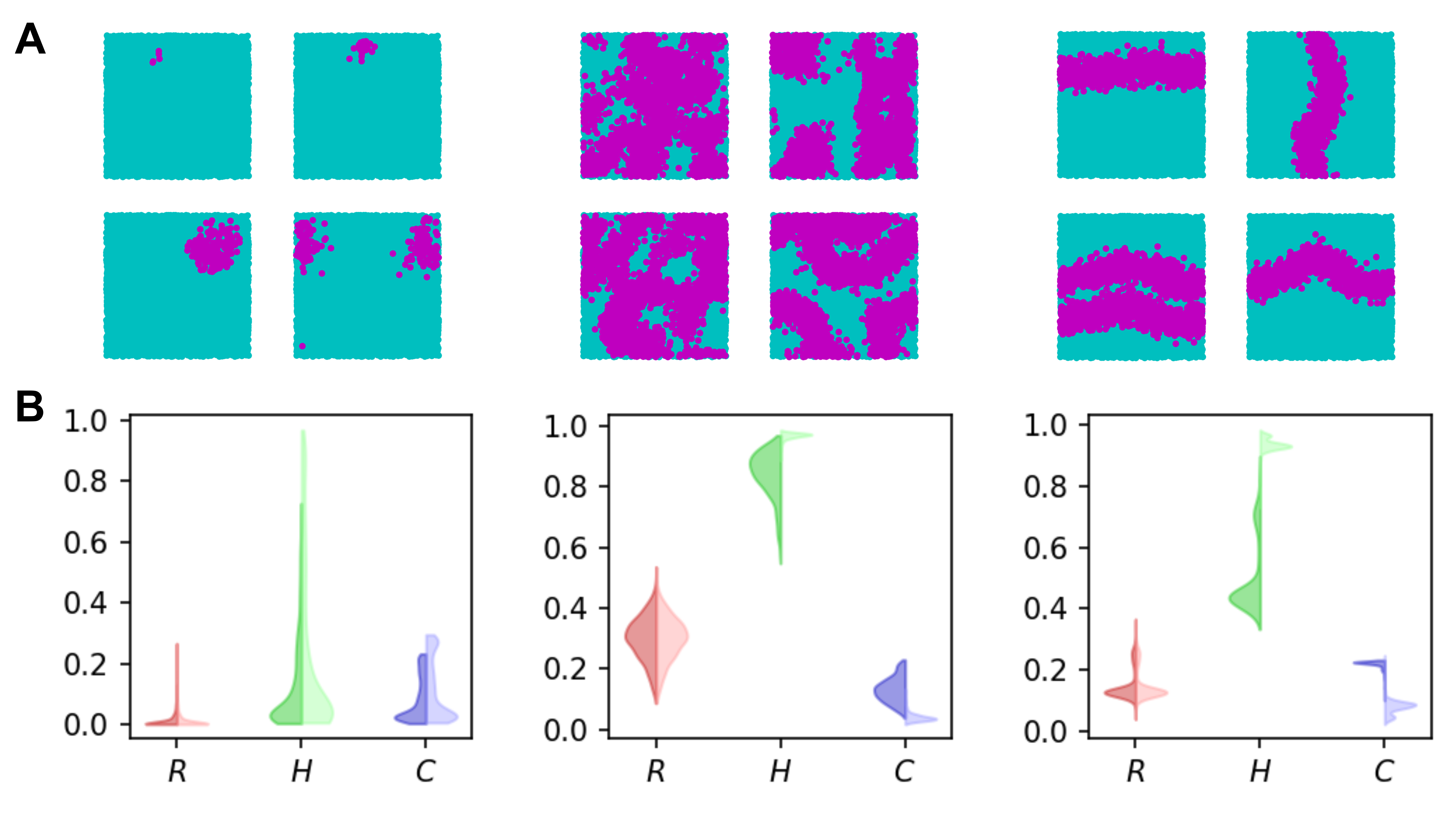}
    \caption{Examples of spatial triadic percolation patterns and their properties. 
    \textbf{A} Examples of the triadic percolation patterns: Small clusters (first column),  Octopus (second column), and Stripes (third column). Active nodes are highlighted in pink and inactive nodes are in turquoise. 
     \textbf{B} Quantitative characterizations of the triadic percolation patterns. The size $R$, permutation entropy $H$, and complexity $C$ of patterns are shown to reveal important quantitative differences between patterns. For each pattern, a surrogate random pattern was created with the same size $R$ and randomly selected active nodes. The distributions of $R$, $H$, $C$ for each pattern class are shown by two-sided violin plots, where the left-hand side corresponds to the spatial triadic percolation patterns, and the right-hand side to the surrogate random patterns. 
    The values of the entropy $H$ and the complexity  $C$ are obtained by embedding active nodes into a $30 \times 30$ grid with sliding partitions of size $d_x=2$ and $d_y=2$ (see Methods section for details).
    The spatial network with triadic interactions is  obtained by a realization of the model with parameter values $N=10^4,c^+=c^-=0.2$, $d_r=d_0=0.25$, $c=0.4$.
    The triadic percolation patterns are derived (for classification details see Fig. $\ref{fig:figure2_TDAmethod}$) from simulations with $p=0.1,\ 0.2,\ .... 1.0$, each including $500$ time-steps (after a transient period of $500$ steps).
    }     
    \label{fig:figure3_patterns}
\end{figure}

\subsection{Information theory of triadic percolation patterns}
The emergent patterns can be further investigated  with information theory tools.  
In particular, for each class of triadic percolation patterns we study the distribution of the size $R$ of the giant component, together with the permutation entropy $H$ and the complexity $C$ of the patterns.
While $R$ is the usual order parameter for percolation, the permutation entropy  $H$ and the complexity $C$ were formulated originally for quantifying local spatial patterns on 2-dimensional images \cite{sigaki2018history} and patterns \cite{bandt2002permutation, ribeiro2012complexity}.

Permutation entropy $H$ quantifies the ``randomness" of the local spatial patterns by calculating the entropy of pixel permutations. Higher permutation entropy values indicate greater irregularity in the data. 
Complexity $C$ provides additional information about the degree of correlational structure by considering larger spatial patterns of motifs. A high permutation complexity value indicates the presence of diverse, less repetitive patterns of the data. Complexity will reach a minimum on either completely random patterns or regular patterns, and a maximum on patterns with ``hidden" structure \cite{rosso2007distinguishing, martin2003statistical, martin2006generalized}.

In Fig. $\ref{fig:figure3_patterns}$B we report the distributions of $R$, $H$ and $C$ for each pattern class. As already hinted by qualitative observations, the three  triadic percolation pattern classes differ on the average size $\bar{R}$,  with clusters being typically small, octopus spanning up to half of the nodes of the network, and stripes presenting intermediate values. 
 However $R$ is not enough  to distinguish individual patterns, as evidenced by the overlap in the distribution of $R$ values between pattern classes.  
For instance, a stripe pattern may include more than one stripe, resulting in a stripe pattern with a relative large size, and similarly cluster patterns may be relatively large, i.e. of similar size to the typical stripes (for instance, see Fig. $\ref{fig:figure3_patterns}$B third column). 

Quantitatively, the topology or the complexity of the pattern can neither be distinguished by means of the size $R$ alone.
In particular, random patterns with no spatial structure can emerge with any given value of $R$ (on non-spatial networks). 
On the contrary, the permutation entropy $H$ and complexity $C$ provide extra information on the rich spatial organization of the patterns, which is remarkably different from that of random patterns. 
To quantify this finding, in Fig. $\ref{fig:figure3_patterns}$B we compare the actual distribution of $H$ and $C$ (left-hand side of each violin plot, darker color) with the distribution obtained from surrogate random patterns with the same size as each of the spatial triadic percolation patterns (right-hand side of each violin plot, darker color). 
As it can be observed, octopus and stripes are vastly different from random patterns, with a) consistently lower permutation entropy $H$, b) higher complexity  $C$,  c) wider distributions of $H$ and, in the case of octopus, also wider distribution of $C$.
Cluster patterns on the contrary are less different from surrogate random patterns, suggesting that the observed values of $H$ and $C$ are mostly driven by the pattern size $R$ in the case of uniform clusters.

The three classes of triadic percolation patterns differ in terms of absolute values of entropy and complexity. 
Octopus patterns show on average the largest permutation entropy, followed by stripes, whereas clusters in general present low values of $H$. 
Regarding complexity, this is highest for stripes, followed by octopus, whereas clusters present typically small values but with large deviations.
Thus, on this meso-scale description, stripes are the least random pattern.

Finally, we notice that not only the average value of $H$ and $C$ differs between pattern classes, but their within-class distributions also show specific properties.
Octopus show the largest variation of $H$ and $C$, evidencing large within-class variability. For stripes, two well-defined maxima of $H$ are observed, corresponding respectively to single and double stripes. Remarkably, these barely differ in terms of complexity, which is in strong contrast with surrogate random patterns of the same size, for which the two maxima are observed again. Thus all stripes are similar in terms of complexity, whereas the variability in octopus does lead to different complexity.

\begin{figure}[!tbh]
    \centering \linespread{1.1} \selectfont
    \includegraphics[width=\columnwidth]{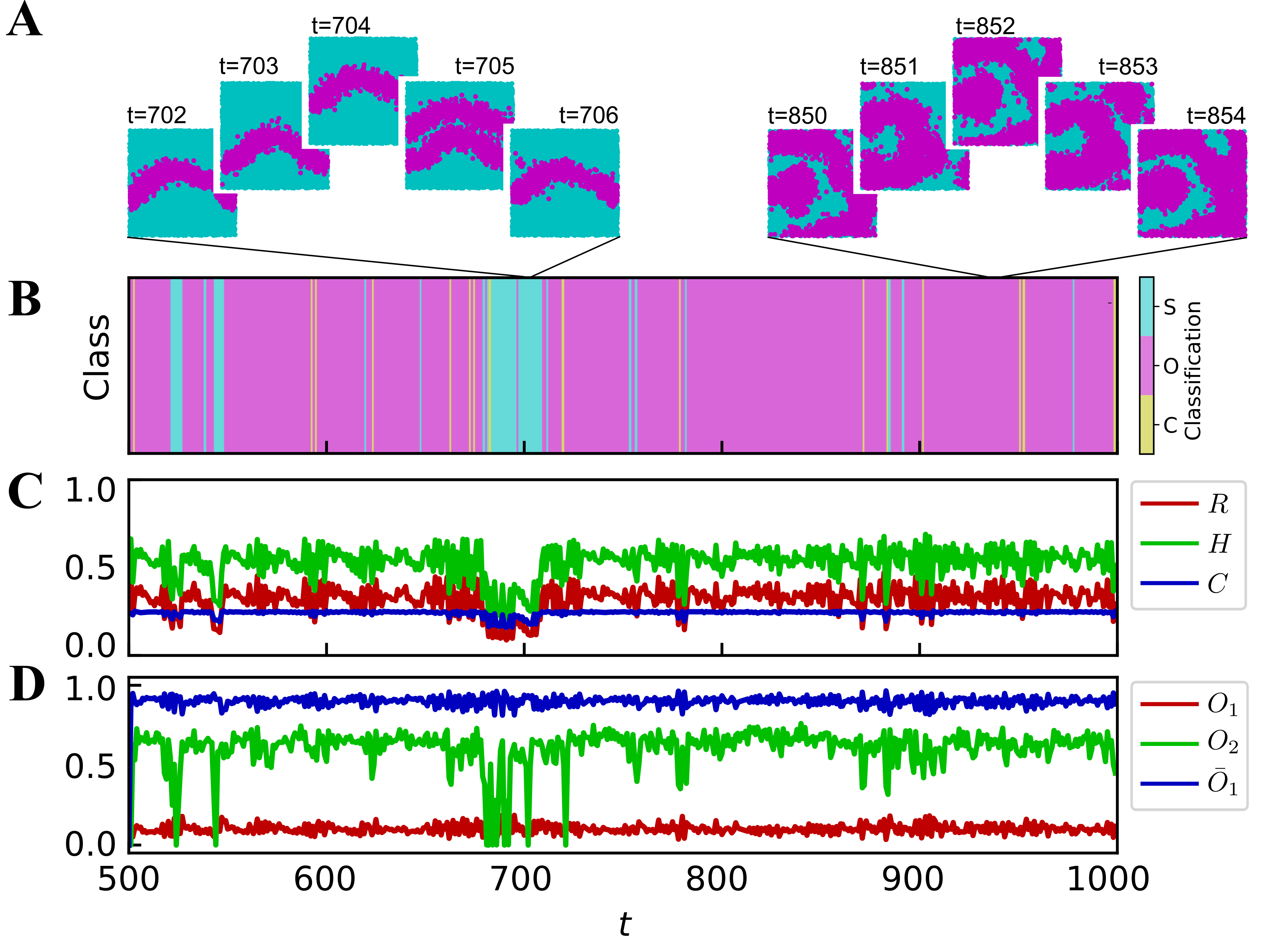}
    \caption{
     { Exemplary intermittent time-series of spatial triadic percolation patterns.} 
     \textbf{A-D} Time-series of triadic percolation patterns observed for $p=0.7$ on   the same network realization as Fig. $\ref{fig:figure2_TDAmethod}$.  
     Panel A shows the spatial triadic percolation patterns at exemplary time steps. 
     Panels B and C show respectively the pattern classification (cyan for stripes S, pink for octopus O, and yellow for clusters C),  the size of the giant component  $R$,  the entropy $H$ and  the complexity $C$ of the triadic percolation patterns a a function of time.  
     To illustrate the temporal dynamics of the triadic percolation patterns (panel D), we monitor the overlap $O_{\tau}$ between active patterns at  lag  $\tau=1,2$ (leading to $O_1$ and $O_2$), and the overlap $\bar{O}_1$ between the active and inactive patterns at lag $\tau=1$.  
    }
    \label{fig:figure4_timeseries_RHC_TDA}
\end{figure}

\section{Dynamics of triadic percolation patterns}

Triadic percolation displays a non-trivial temporal organization of the dynamics of its emergent topological patterns. 
The topology of the triadic percolation  patterns displays metastability, with patterns belonging to the same topological class persisting for varying lifetimes depending on both the quenched and the annealed disorder of the triadic percolation model. 
In particular, intermittent time-series of the triadic percolation patterns occur for a wide range of parameters values.
For instance, in  Fig. $\ref{fig:figure4_timeseries_RHC_TDA}$ we show a time-series in which both octopus and stripes occur in an intermittent and dynamic fashion. 
Videos of  exemplary time-series of triadic percolation patterns are included in the Supplementary Videos. 

The temporal behavior of the fraction of nodes in the giant component $R$, and of the information theory measures (entropy $H$ and complexity $C$) capture important aspects of the dynamics of triadic percolation (see Fig. $\ref{fig:figure4_timeseries_RHC_TDA}$C). 
However there are several non-trivial spatio-temporal effects which go beyond the information that can be extracted from $R$, $H$ and $C$ alone.

The emergence of time-varying triadic percolation patterns is a direct consequence of local connectivity and also local positive and negative regulation. 
Local negative regulation leads to emergent self-inhibition in the macroscopic scale, whereas local positive regulation implies that activity can only emerge on the neighbourhood of previously active regions.
Together, these mechanisms create and effect surface tension, and give rise to the emergent macroscopic patterns of different shapes, see Supplementary Information (SI) for further details.  
As a consequence, we observe that in the deterministic case stripes always emerge through surface minimization. 
Thus on a torus constructed from a rectangular map (rectangular torus), stripe patterns only occur along the shorter dimension of the rectangular torus (see SI).

Crucially, this mechanism  induces an intrinsically dynamic nature of the triadic percolation patterns $\bm{s}(t)$. 
Moreover, given that positive regulation occurs  on a local scale, the active pattern at time $t+1$, $\bm{s}(t+1)$, can only appear on the neighbourhood of the active pattern at time $t$, $\bm{s}(t)$. 
In combination with effective self-inhibition, this causes the tubular geometry of the stripe and octopus patterns.

\begin{figure}[!tbh]
    \centering \linespread{1.1} \selectfont
    \includegraphics[width=0.9\columnwidth]{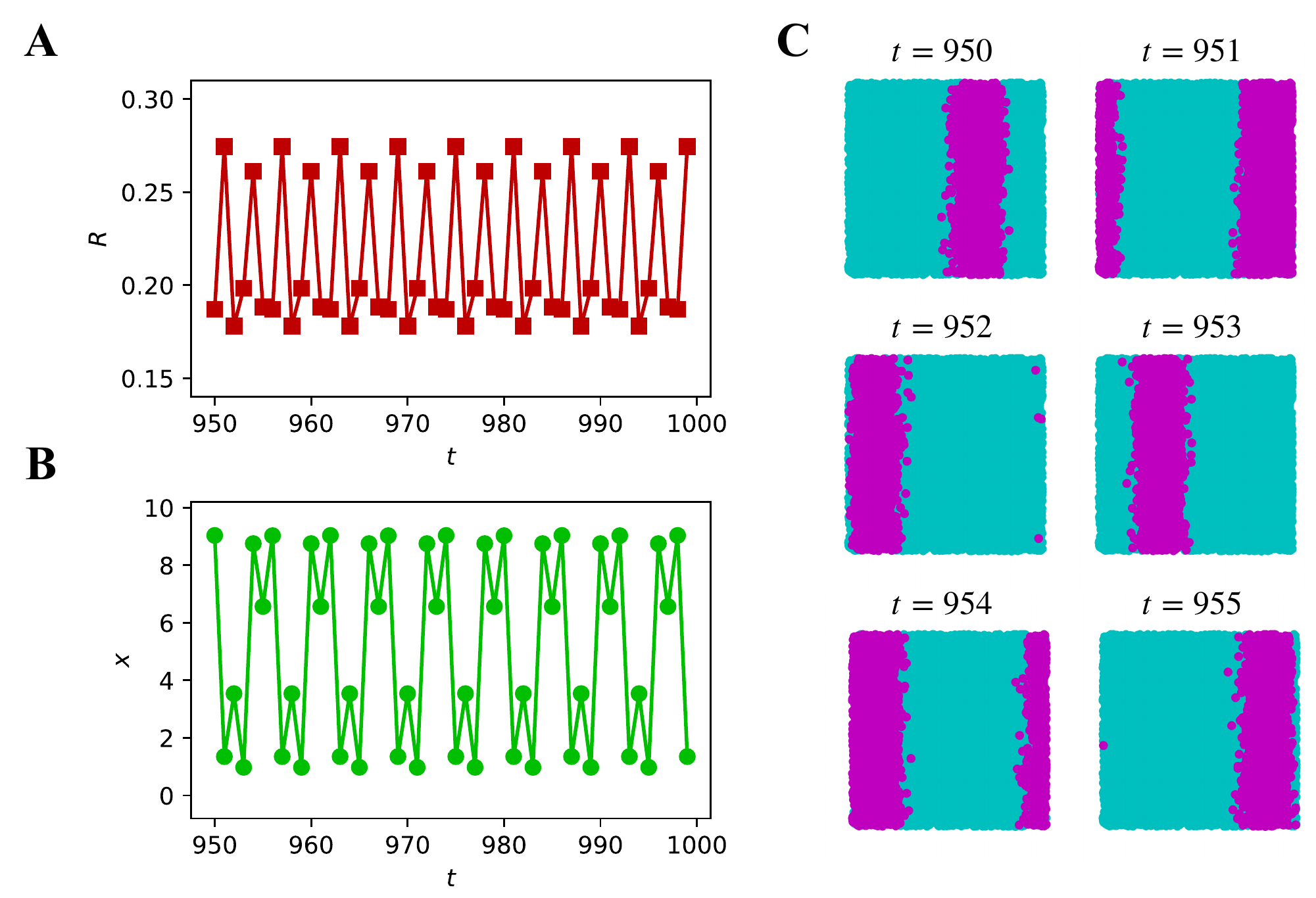}
    \caption{{Blinking of stripe patterns.} \textbf{A} Relative size of the giant component $R$ is shown a function of time $t$. 
    \textbf{B} The $x$-coordinate of the barycenter of the patterns is shown as a function of time $t$.  
    \textbf{C} Examples of stripe patterns. Active nodes are highlighted in pink and inactive nodes are in turquoise. All these measures clearly indicate the sustained blinking of the triadic percolation patterns, which in this case is a periodic dynamics with period $6$. 
    The spatial network is formed by $N=10^4$ nodes. The parameters for generating the spatial  network with triadic interactions are  $c^+=0.4$, $c^-=0.2$, $d_r=d_0=0.25$, $c=0.4$, $p=1$. 
    }
    \label{fig:figure5_blinking_stripes}
\end{figure}

The combination of self-inhibition and border-excitation also induces the temporal dynamics of the patterns, characterized by two dominant mechanisms at short time scales: short-time-blinking (ST-blinking) and diffusion. 
ST-blinking consists of the alternation of complementary patterns such that $\bm{s}(t+1)\approx 1-\bm{s}(t)$.
ST-blinking occurs for large patterns (typically octopus) such that the pattern and its neighbourhood encompass the whole structural network. 
Diffusion, on the contrary, occurs for smaller patterns  (small clusters) whose neighbourhood does not cover all the inactive region. 
In this case the active pattern may not return after two steps, as more options are available, the probability of each one depending on the specific local fluctuations of connectivity.

ST-blinking and diffusion can be quantified by the overlap between patterns at different time-steps, defined as $O_\tau=\sum_i s_i(t)s_i(t+\tau) / \sum_i s_i(t)$, where $O_\tau(t)\simeq 1$ indicates that patterns $\bm s(t)$ and $\bm s(t+\tau)$ are similar, whereas values close to $0$ indicate that they are significantly different.  
In particular, in Fig. $\ref{fig:figure4_timeseries_RHC_TDA}$D we show the overlap between consecutive active patterns, $O_1=O_1(t)$, and between patterns separated by two steps, $O_2=O_2(t)$, and the overlap between the active pattern at time $t$ and the inactive pattern at time $t+1$, $\bar{O}_1=\sum_i s_i(t) (1-s_i(t+1)) / \sum_i s_i(t)$. 
Due to emergent self-inhibition, $O_1$ is always low and $\bar{O}_1$ high, whereas $O_2$
distinguishes between ST-blinking and diffusion: it is large for ST-blinking, and small for diffusion. 
In the exemplary case of Fig. $\ref{fig:figure4_timeseries_RHC_TDA}$, we observe in general high $O_2$, indicating a domination of ST-blinking, although it intertwines with diffusion when stripes emerge (see panels A,D around $t=700$).
In general, we have found that octopus patterns blink on short time-scales, and clusters diffuse, whereas stripes show a mixed behavior, as shown in Supplementary Figs. $\ref{fig:sup_fig_overlap}$ and $\ref{fig:sup_fig_barycenter}$.

The microscopic organization of the structural and regulatory networks shapes the dynamics of the patterns, as easily observed for stripes.
Assuming fully homogeneous local connectivity, $1$D diffusion (akin to a random walker on a $1$D lattice) would be the expected behavior for stripes: a horizontal stripe may move up or down, with equal probability.
However, we observe a location-dependent asymmetry in the movement, that effectively traps the stripes into preferred locations. 
This asymmetry is induced by microscopic inhomogeneities or quenched disorder. 
The effect of quenched disorder is less evident for increased random  damage, which adds {annealed} noise into the dynamics (see Supplementary Fig. $\ref{fig:sup_fig_barycenter}$). 

Finally, we note that triadic percolation patterns can blink, i.e. display sustained periodic dynamics, depending on the properties of the structural and regulatory networks (see Fig. $\ref{fig:figure5_blinking_stripes}$). 
Our previous discussion accounted for the case of $c^+=c^-$. 
For illustration purposes we now consider the case of strong positive regulation ($c^+=2c^-$) and the deterministic scenario (no random damage, $p=1$). 
In this case, stripes are the only observed pattern (as for $c^+=c^-$, $p=1$). 
For instance, in  Fig. $\ref{fig:figure5_blinking_stripes}$A,B we show evidence of  blinking with a period $6$ dynamics for both $R$ and the barycenter location $x$. This blinking behavior consisting of sustained periodic dynamics of the triadic percolation patterns is also validated by the inspection of the actual triadic percolation patterns shown in Fig. $\ref{fig:figure5_blinking_stripes}$C.

These results allow us to draw a general interpretation of the observed spatio-temporal phenomenology.
Triadic percolation on spatial networks leads to complex spatio-temporal dynamics. 
Two spatio-temporal scales emerge. 
Spatially, the patterns are heterogeneous, with the emergence of well-defined active and inactive regions. 
At short time-scales we observe a  dynamics dominated by ST-blinking, for intermediate random damage and large patterns, and by diffusion, for large random damage and small (clusters) patterns. 
In the case of stripes, quenched disorder caused by random local connectivity breaks translational symmetry, and preferred locations for stripes emerge at the mesoscopic scale, leading to ST-blinking and in some cases to sustained periodic blinking of triadic percolation patterns, particularly for the deterministic (no random damage) case. 
The specific nodes becoming active on each appearance of the macroscopic stripe may vary and the barycenter shows small microscopic variations, but it remains contained within the width of a macroscopic stripe. 
In absence of sustained periodic blinking, a  combination of random (annealed) damage and quenched disorder, caused by random local connectivity, deforms the emergent patterns on large temporal scales. 

\begin{figure}[!tbh]
    \centering \linespread{1.1} \selectfont
    \includegraphics[width=0.3\columnwidth]{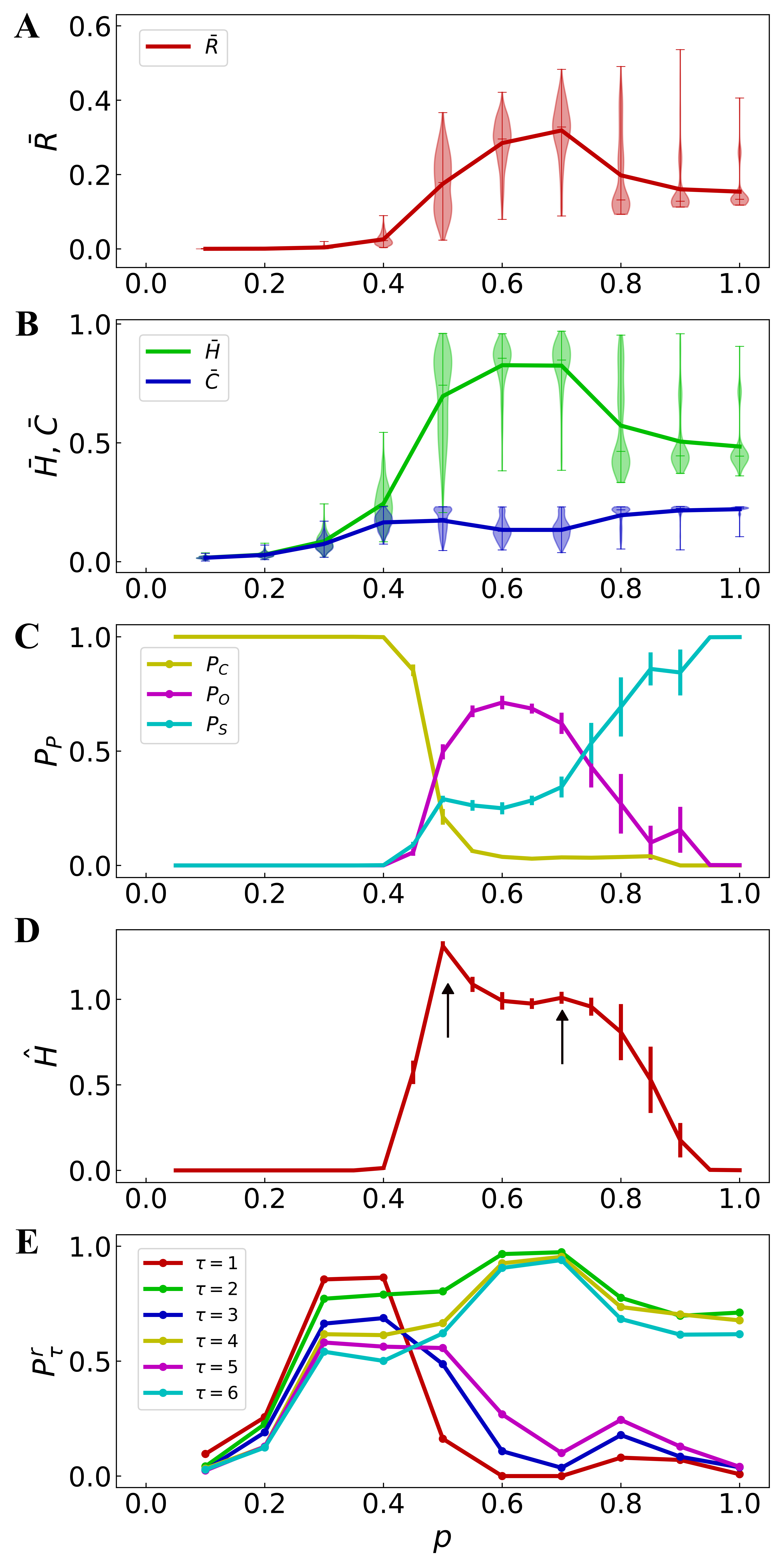}
    \caption{Phase diagram of spatial triadic percolation.
   Panels \textbf{A-B} The average size  $\bar{R}$ of the giant component (panel \textbf{A}),  the average values of the permutation entropy $\bar{H}$, and the complexity $\bar{C}$ of the triadic percolation patterns (panel \textbf{B}) are plotted as a function of $p$. Panels \textbf{C-D} The TDA classification is adopted to enrich the phase diagram with static and dynamical information about the topology of the patterns. The fraction of triadic percolation patterns classified as Clusters ($P_C$), Octopus ($P_O$) and Stripes ($P_S$)  (panel \textbf{C}) and the entropy rate $\hat{H}$ of the pattern time-series (panel \textbf{D}) are plotted as a function of $p$. The black arrows indicate two local minima. Panel \textbf{E} displays the  probability  $P^r_\tau$ of return after $\tau$ steps as a function of $p$, averaged over the time-series, revealing the nature of the short time dynamics of the triadic percolation patterns. 
   The spatial network with triadic interaction is formed by $N=10^4$ nodes and generated with parameter values  $c^+=c^-=0.2$, $d_r=d_0=0.25$, $c=0.4$ in all panels.  Panels A,B,E are obtained from the same data corresponding to $500$ steps of the triadic percolation dynamics. The entropy $\bar{H}$ and the complexity $\bar{C}$ are obtained by embedding active nodes into a $30 \times 30$ grid with sliding partitions of size $d_x=2$ and $d_y=2$.   The accompanying violin plots show the distribution of values of each variable. Panels C and D are obtained from the average of  $10$ different iterations of the triadic percolation dynamics, each including $1500$ steps of the dynamics. The curves show the average values, and the errorbars the standard deviation among network realizations. Data were considered after a transient of $500$ steps in all cases.
    }
     \label{fig:figure6}
\end{figure}

\section{Phase diagram of spatial triadic percolation}

The phase diagram of triadic percolation, studied as a function of the control parameter $p$ (see Fig. $\ref{fig:figure6}$) reflects its complex spatio-temporal dynamics.  
In order to investigate this phase diagram, we first monitor the relative size of the giant component $R$ as a function of $p$ (see Fig. $\ref{fig:figure6}$A). 
The first important observation that we make is that  in average, the relative  size of the giant component $\bar{R}$ displays a maximum and is hence not monotonically increasing with $p$ as in standard percolation. Moreover, the relative size of the giant component is roughly bounded by half of the nodes of the network. 
This is  a consequence of the competition between 
local positive and negative regulation,  given that active nodes at time $t+1$ are limited to the area nearby (but, crucially, excluding) the active pattern at time $t$. As we have discussed extensively in the previous sections, however, the value $R$ does not provide detailed information on the triadic percolation patterns. 
To investigate the  nature of the triadic percolation patterns, we monitor the average of  the information theory measures $H$ and $C$ (given by $\bar{H}$ and $\bar{C}$) as a function of $p$ (see Fig. $\ref{fig:figure6}$B). 
Most relevantly we find that also the average permutation entropy $\bar{H}$, has a maximum for intermediate values of $p$ indicating more random patterns. 
The quantities $R, H$ and $C$ have the advantage of being unsupervised quantities, note however that, due to their large within-class variability  (as reported in Fig. $\ref{fig:figure3_patterns}$),  $R, H,C$ have limited ability  to detect the regions of the phase diagram with distinct topological organization (see Fig. $\ref{fig:figure6}$A, B).
The phase diagram of triadic percolation is therefore enriched by considering the results of TDA analysis (see Fig. $\ref{fig:figure6}$C).
The TDA analysis reveals three regions of the phase diagram, each dominated by a single type of pattern, although values of $p$ with significant coexistence of patterns are observed.
For small values of $p$ (large probability of random edge damage), low-activity patterns, i.e. clusters, are most likely.
On the other hand, for large values of $p\simeq 1$ (strongly suppressed random edge failure), most regulation links are active, 
and the combination of local positive and negative regulation results in stripe patterns due to the border-minimization effect (see SI). For intermediate values of $p$  the octopus topology emerges as the most likely. 
The three TDA-detected regions are observed for ranges of $p$ values that depend on the other parameters of the model, although octopus and stripe patterns require that the active cluster is percolating, ensuring the connectivity of the giant component. Moreover, stripes always emerge for $p=1$ (no random damage).

As triadic percolation is a dynamical process, the phase diagram is not complete if we do not provide information about  the dynamical nature of the  time-series of patterns as a function of the control parameter $p$.
To this end,  we have considered two quantities that capture the temporal dynamics of the macroscopic patterns: the entropy rate $\hat{H}$ of the pattern time-series, and the return probability $P^r_{\tau}$ of the overlap $O_\tau$ time-series defined in Sec. $3$. 
These quantities  probe  the temporal dynamics of the system at different scales:  the entropy rate $\hat{H}$ captures the long-term behavior (as revealed  by Fig. $\ref{fig:figure5_blinking_stripes}$) and $P^r_{\tau}$ quantifies the short time dynamics typically observed,
and can be used to classify diffusion and ST-blinking of patterns in time.

The entropy rate $\hat{H}$ (see Methods for details) quantifies the information content of the time-series of patterns as derived from the TDA classification analysis (in fig. $\ref{fig:figure6}$C).  
Therefore the complex triadic percolation dynamics is codified by the single time-series of the pattern classification (a series of letters with letters indicating the three distinct type of topological patterns --clusters, octopus and stripes).  
Very predictable time-series have low entropy rate, highly unpredictable ones have high entropy rate. 
Hence the entropy rate $\hat{H}$ can be used to probe intermittent time-series of patterns and can be related to  the complexity of the  process \cite{amigo2004estimating} (see SI for details). 
Interestingly the entropy rate $\hat{H}$ presents evidence for a dynamical phase transition  at $p=p_c\simeq 0.4$ (see Figure $\ref{fig:figure6}$D).
For $p<p_c$ only one type of pattern is observed (clusters)  and the entropy rate is zero, i.e. $\hat{H}=0$; beyond this phase transition different types of patterns coexist and thus we have $\hat{H}>0$, while for $p\simeq 1$ the entropy  rate is again zero, i.e. $\hat{H}=0$ and only stripe patterns are observed.
Moreover the entropy rate $\hat{H}$ displays two local maxima corresponding to values of $p$ with more diverse compositions of pattern types. 
The first local maximum  (smaller $p$) is higher due to the coexistence of the three pattern-classes (clusters, octopus and stripes) while in correspondence of the second local maximum  (larger values of $p$) we observe coexistence only of two pattern-classes (octopus and stripes).

The second measure we introduce is the return probability $P^r_{\tau}$ that measures the probability that a pattern re-occurs (or returns) after $\tau$ steps, and is analogous to the return probability on a diffusion problem  \cite{millan2021local}.  
Hence this measure differs from the first not only because it probes the dynamics at smaller time-scales, but also because it is very sensible to the geometrical details of the patterns (typically having the same topological classification).
In this framework, $P^r_{\tau}$ is measured by binarizing and averaging over time the corresponding overlap metric $O_\tau$
(see Methods section for details).
As observed in Fig. $\ref{fig:figure6}$E,
$P^r_{\tau}$ follows a non-trivial trend in the three regions of the phase diagram dominated by the three different topologies (panels A).
For even $\tau$ values, there is a maximum in the octopus-dominated region (intermediate $p$ values), with sustained high values in the stripes region (large $p$ values) and minimum values for the cluster-dominated (small $p$ values) region. 
For odd $\tau$ values we find low values on the octopus and stripes region due to the strong ST-blinking effect.
Thus, octopus patterns and stripes have a larger probability of re-occurrence at short time-scales (associated with a ST-blinking dynamics), than clusters. 
In absence of sustained periodic blinking,  by increasing $\tau$ we explore longer time-scales in the system and we observe that the return probability $P^r_{\tau}$ decreases with $\tau$ in all cases, evidencing that eventually the macroscopic patterns are deformed at longer time-scales, as a consequence of the interplay between quenched disorder and random damage. 

The phase diagrams derived from the temporal dynamics ($\hat{H}$ and $P^r_\tau$) follow closely those derived from the spatial organization (pattern occurrence, $\bar{R}$, $\bar{H}$, $\bar{C}$), indicating that each pattern class not only presents distinct geometry and homology but also a characteristic macroscopic evolution of the patterns.

As we have seen, the phase diagram of triadic percolation provides a comprehensive understanding of  triadic percolation. 
In particular, this phase diagram covers  the topology, the information content and the dynamics of this very rich dynamical process going beyond a description of triadic percolation based exclusively on the size of the giant component.

\section{Conclusions}

In summary, triadic interactions are fundamental higher-order interactions present in a variety of complex systems, ranging from brain networks to ecosystems, that can dramatically change the properties of percolation, as captured by triadic percolation. 
Here we show that, on spatial networks,  triadic percolation leads to the emergence of a  time-varying giant component that displays topologically distinct patterns.  
In order to investigate the spatio-temporal properties of triadic percolation we combine network science, TDA, information theory and  the theory of nonlinear dynamical systems to describe this remarkable critical phenomenon. 
The giant component displays  patterns that  are classified through persistent homology into three different classes:  small clusters (formed by a localized scattered set of points),  octopus patterns (patterns with non-trivial persistent diagram), and stripes (patterns of points going around the torus). 
These emergent spatial triadic percolation patterns are remarkably different from random patterns, as quantified by our information theory analysis, which reveals the lower entropy and larger complexity of these patterns with respect to random unstructured patterns.  
These patterns have a very non-trivial dynamics revealing that the giant component of triadic percolation has a topology that  can change significantly over time. 
We show that for some parameter values the time-series of patterns displays intermittency between different topological classes. 
Moreover, in the case of stripes we provide evidence of blinking behavior with the barycenter of the giant component oscillating periodically. 
These findings are summarized using a phase diagram of triadic percolation, indicating the regions where patterns of a given type are more likely to occur. 
The phase diagram also shows how statistical, information theory, and temporal observables of the complex spatio-temporal dynamics of the giant component change as a function of the control parameter leading to a comprehensive understanding of this complex dynamics. 

The observed spatio-temporal modulations of the topology and geometry of the giant component open new perspectives in percolation theory and its applications.
As giant components changing dynamically in time are observed in a large variety of real systems, these  findings have the potential to transform our theoretical understanding of these systems, with  significant potential applications in neuroscience.

\section*{Material and Methods}
\subsection{Spatial higher-order networks with triadic interactions}

We consider spatial higher-order networks with triadic interactions embedded in a $2D$ square of size $L$ with periodic boundary conditions (a torus). 
The density of nodes is indicated by $\rho$, with $\rho$ fixed as $100$ nodes per unit square in all simulations of this paper. 
The structural networks $G_s=(V,E)$ contain edges drawn randomly between each pair of nodes with probabilities decaying exponentially with their Euclidean distances. 
Specifically, a pair of nodes $i$ and $j$ are connected with probability
\bea
P_{ij} = c e^{-d_{ij}/d_0}
\eea
where $d_{ij}$ denotes the Euclidean distance between nodes $i$ and $j$, and $d_0$ denotes a typical length of structural links. 
The average structural degree $\avg{k}$ is controlled by $0<c\leq 1$.
The spatial regulatory network is generated as follows.
First, we define the coordinate of a structural link $\ell$ as the midpoint of its two end-nodes.
Then, a positive or a negative regulatory interaction between a  node $i$ and a structural links  $\ell$ is drawn  with probability  $\hat{P}^{+}_{i\ell}$ and $\hat{P}^{-}_{i\ell}$ respectively.  The probabilities $\hat{P}^{+}_{i\ell}$ and $\hat{P}^{-}_{i\ell}$ are given by
\bea
\hat{P}^+_{i \ell} = c^+ e^{-d_{i\ell}/d_r},\nonumber \\
\hat{P}^-_{i\ell} = c^- e^{-d_{i\ell}/d_r},
\label{Ppm}
\eea
where we exclude the possibility of conflicting regulatory interactions.
In other words, we impose that if a node is a positive regulator of a link it cannot be simultaneously a negative regulator of the link.
In Eq.($\ref{Ppm}$) $d_r$ defines the typical length of regulatory interactions, and $c^+$ and $c^-$ (with $c^+>0,c^->0$ and $c^++c^-\leq 1$) control the average number of nodes that regulate a link positively  or negatively.

For $d_r \to \infty$ and $d_0 \to \infty$, the spatial network with triadic interactions reduces to the Erdös-Renyi network with random triadic interactions that is defined and studied in Ref. \cite{sun2023dynamic}. 

\subsection{TDA based classification}
The spatial triadic percolation patterns are here  classified by means of topological data analysis (TDA). 
TDA allows us to identify a pattern's shape and its invariant topological properties, with moderate noise tolerance \cite{blevins2020reorderability, blevins2021variability, zomorodian2012topological, zomorodian2005topology}.
The TDA  of a set of points proceeds by first generating a simplicial complex that represents the data at different values of a \emph{filtration} or threshold parameter $f_s$, and then evaluating the homology classes of the simplicial complex as function of $f_s$.  
A simplicial complex is a finite collection of simplices $K$ such that i) every face of a simplex in $K$ also belongs to $K$, and ii) for any two simplices $\sigma_1$ and $\sigma_2 \in K$, if $\sigma_1\cap \sigma_2 \neq \varnothing$, then $\sigma_1 \cap \sigma_2$ is a common face of both $\sigma_1$ and $\sigma_2$. 
A $d$-simplex is the convex hull of $d+1$ points: a $0-$simplex is a point, a $1-$simplex is an edge, a $2-$simplex is a triangle and so on.
Here we consider the Vietoris-Rips filtration method to build simplicial complexes that represent the data. 
Given a pattern of active nodes at time $t$, $\bm{s}(t)$, the corresponding Vietoris-Rips complex is the filtered complex $VR_s(\bm{s}(t))$ that includes all $\delta-$simplices, $\delta\leq d$, such that all pairwise distances between the nodes in the simplex are equal or less than $f_s$, that is $VR_s(\bm{s})=\left\lbrace \left[ v_0,...,v_n \right]  \forall i,j\ d(v_i,v_j)\leq f_s \right\rbrace$. 
The $k$-homological classes \cite{zomorodian2012topological, centeno2022hands} of a  simplical complex are in one-to-one correspondence with its  independent $k-$dimensional holes: in dimension $0$, these are connected components, in dimension $1$, cycles (also called loops), in dimension $2$, $2$-dimensional holes (like in a triangulated sphere), and so on.
Persistent homology tracks the  homology classes as the filtration parameter $f_s$ increases and detects which topological features persist across different scales \cite{aktas2019persistence}.
The filtration values at which each homological class emerges (birth) and disappears (death) can be recorded on a persistence diagram PD (showing the death value as a function of the birth value) and characterize a given shape or pattern. Points further from the diagonal mark features that survive for long filtration intervals.
PDs of different patterns can be compared by measuring the distance between them. Here we considered the Wasserstein distance, which matches pairs of points between the two diagrams and measures the $L^p$ distance between them. Points that cannot be matched to a point in the other diagram are matched to the diagonal. 

The patterns were classified into Clusters (C), Octopus (O) and Stripes (S) using persistence homology. First, a set of representative template patterns ($33$ for each pattern class) were manually identified (an example for each class is given in the insets of Fig. $\ref{fig:figure2_TDAmethod}$A), and PDs were obtained for each one (Fig. $\ref{fig:figure2_TDAmethod}$A). 
Secondly, persistence homology was applied to each pattern $\bm{s}(t)$ of a given simulation of the system. 
The Wasserstein distance between the PD of $\bm{s}(t)$ and all template PDs was measured, and state $\bm{s}(t)$ was assigned the pattern class of the closest template (Fig. $\ref{fig:figure2_TDAmethod}$D).
For the VR filtration we consider simplices up to $d=2$, and the persistence diagrams included the $0$- ($H_0$) and $1$- ($H_1$) dimensional holes. The total Wasserstein distance was defined as the sum of the $H_0$- and $H_1$-associated distances.
All TDA analyses were performed using the giotto-tda python library.

\subsection{Entropy and Complexity}
The structural and information theory properties of the spatial triadic percolation patterns include their permutation entropy and statistical complexity. The measure of permutation entropy was first proposed by Bandt and Pompe \cite{bandt2002permutation} to measure the complexity of one-dimensional time-series. The approach was generalized to higher-dimensional data and an additional measure of complexity called López-Ruiz-Mancini-Calbet (LMC) complexity was proposed. The LMC complexity provides structural information that is not included in the entropy measure \cite{lopez1995statistical, ribeiro2012complexity}. 
The permutation entropy is calculated based on the permutation of local partitions. Consider a two-dimensional pattern that is represented by a matrix. We consider all local partitions of size $(d_x, d_y)$, \ie, submatrices of size $d_x$ by $d_y$, where $d_x$ and $d_y$ are called embedding dimensions \cite{sigaki2018history}. Let us consider a simple case where $d_x=d_y=2$. Thus each submatrix can be written as
\bea
A = 
\begin{bmatrix}
a_0 & a_1\\
a_2 & a_3
\end{bmatrix}
\eea
Reshaping the submatrices to one-dimensional vectors, we can categorize them into different ordinal patterns. For instance, pattern $\pi_1=(0,1,2,3)$ denotes all submatrices in which $a_0<a_1 < a_2<a_3$ and pattern $\pi_2=(1,0,2,3)$ denotes all submatrices in which $a_1<a_0<a_2<a_3$. There are in total $(d_x d_y)!=24$ ordinal patterns in this example. Thus we can define the distribution of ordinal patterns with given embedding dimensions $d_x$ and $d_y$. The probability $P(\pi)$ of having an ordinal pattern $\pi$ is calculated by
\bea
P(\pi) = \frac{\text{number of submatrices that have pattern $\pi$}}{\text{total number of submatrices of size $d_x \times d_y$}}
\eea
and the permutation entropy $S$ is defined as
\bea \label{eq:S_permutation_entropy}
S[P] = -\sum_{\pi} p(\pi) \ln p(\pi).
\eea

The measure $H$ used above is the normalized permutation entropy defined as
\bea \label{eq:H_permutation_entropy}
H[P] = \frac{S[P]}{S_{\max}} = \frac{1}{(d_x d_y)!} S[P].
\eea
The permutation entropy quantifies the amount of ``information" in the patterns. The statistical complexity $C$ composites the measure of information $H$ and the measure of ``disequilibrium" $Q$ \cite{lopez1995statistical}:
\bea
C[P] = Q[P,P_e] H[P].
\eea
The disequilibrium $Q[P, P_e]$ is defined as the extensive Jensen-Shannon divergence that quantifies the distance between the pattern distribution $P(\pi)$ and the uniform distribution $P_e$ \cite{rosso2007distinguishing};
\bea
Q[P,P_e] = \frac{S[(P+P_e)/2]-S[P]/2-S[P_e]/2}{Q_0}
\eea
where $Q_0$ is a normalization constant. Thus, the complexity $C$ will reach zero at the extremes of the ordered pattern ($H=0$) and completely random pattern ($Q=0$). Note that $C$ is not a trivial function and $H$. For a given $H$ there exists a corresponding range of $C$ that provides additional structural information \cite{sigaki2018history, martin2003statistical}.
To calculate the entropy and complexity of patterns formed by active nodes, we transformed the activity pattern with a 2D density distribution over an $M\times M$ grid, over which we measure the permutation entropy $H$ and complexity $C$ of this matrix \cite{pessa2021ordpy}.

\subsection{Temporal analyses}
The temporal dynamics of the emergent dynamics of triadic percolation is quantified by means of the entropy rate $\hat{H}$ of the pattern time-series $x(t)$ derived from the TDA classification. 
That is, $x(t)$ is the categorical time-series indicating the macroscopic state of the system as Cluster (C), Octopus (O) or Stripe (S).
To measure the entropy rate of  a time-series, we consider motifs or words of increasing length $\hat{L}$, and measure the relative count $\tilde{p}_i$ of each word $i$ of such length. 
The entropy rate of $\hat{L}$-words is defined as \cite{amigo2004estimating}:
\begin{equation}
    \hat{H}(\hat{L})=-\frac{1}{\hat{L}}\sum_i\tilde{p}_i \log_2 \tilde{p}_i.
\end{equation}
The  entropy rate $\hat{H}$, is obtained in the limit of infinite word-length, i.e. 
\begin{equation}
    \hat{H} = \lim_{\hat{L}\to \infty} \hat{H}(\hat{L}).
\end{equation}
Here in order to estimate this limit, we take the usual approach of extrapolating the linear trend of $\hat{H}(\hat{L})$ versus $1/\hat{L}$ for $\hat{L}=1,\ 2,\ 4$, as detailed in the Supplementary Information (see Supplementary Fig. $10$.).
Results shown in the main text are averaged over $10$ time-series encoding the topology of $10$ realizations of  the triadic percolation dynamics.

At short-time scales, the dynamics is investigated  by measuring the overlap between patterns at different time-steps, defined as 
\bea
O_\tau(t) = \frac{\sum_i s_i(t)s_i(t+\tau)}{\sum_i s_i(t)}. 
\eea 
$O_\tau(t)\simeq 1$ indicates that the states $\bm s (t)$ and $\bm s(t+\tau)$ are similar, whereas values close to $0$ indicate significantly different states. 
To discriminate whether two states are \emph{macroscopically} equivalent we have set an adaptive threshold $\alpha'$ on $O_\tau$, such that states $t$ and $t+\tau$ are macroscopically equivalent if $O_\tau(t)> \alpha'$ with
\bea
 \alpha'=\frac{\alpha}{T} \sum_{t'} O_{\tau}(t').
\eea
where $\alpha=0.8$ is the baseline threshold. Here $\alpha'$ is  given by the baseline threshold  $\alpha$ rescaled by the average overlap $O_{\tau}$ of the time-series to account for the large fluctuations in the fraction of active nodes in the triadic percolation patterns.
We thereby described each time-series by the binary variable $\tilde{O}_\tau(t)$ indicating whether the patterns at time $t$ and time $t+\tau$ are macroscopically equivalent. 
Note that this analysis disregards the shape of the patterns. 

The variation of the overlap $\tilde{O}_\tau(t)$ in time is measured by the return probability $P^r_\tau$ \cite{millan2021local, millan2021complex}, that indicates the probability that the system returns at time $t+\tau$ to the same state it was at time $t$, namely $P^r_\tau =  \sum_t \tilde{O}_\tau(t)/T$.

\begin{addendum}

 \item  A.P.M. is financially supported by the ``Ram\'on y Cajal'' program of the Spanish Ministry of Science and Innovation (grant RYC2021-031241-I). A.P.M. and J.J.T.  acknowledge the Spanish Ministry and Agencia Estatal de investigación (AEI) through
Project of I+D+i (PID2020-113681GB-I00), financed by MICIN/AEI/10.13039/501100011033 and FEDER ``A way to make Europe'', and the \emph{Consejer\'ia de Conocimiento, Investigaci\'on, Universidad, Junta de Andaluc\'ia} and European
Regional Development Fund (P20-00173) for financial support. 
 
 \item[Author contributions] G.B. designed the research, A.P.M.,H.S., J.J.T., and G.B. performed the research, A.P.M., H.S. and J.J.T. conducted the numerical calculations, A.P.M., H.S., J.J.T. and G.B. wrote the manuscript. H.S. and  A.P.M. contributed equally to this work.
 
 \item[Competing interests] The authors declare that they have no competing interests.

 \item[Correspondence] To whom correspondence should be addressed. E-mail: apmillan@ugr.es 
\end{addendum}

\bibliography{mybib}

\newpage
\begin{center}
	{\Large{\textbf{Supplementary Information \\ ``Triadic percolation induces dynamical topological patterns in higher-order networks "}}}\\ \bigskip
	\large{A.P. Millán, H. Sun, J.J. Torres and G. Bianconi}
\end{center}	
	\setcounter{section}{0}
	\setcounter{table}{0}
	\renewcommand{\thetable}{S\arabic{table}}
\setcounter{equation}{0}
\renewcommand{\theequation}{S\arabic{equation}}

In this Supplementary Information we provide additional material demonstrating the mechanisms for the emergence of short-time-blinking (ST-blinking) and diffusion of the triadic percolation patterns.
Note that the results discussed here are relative to the extensive region of the parameter space in which sustained periodic blinking is not observed,  such as for  $c^{+}=c^{-}$.

\renewcommand{\figurename}{Supplementary Fig.}


\begin{figure}[!tbh]
    \centering \linespread{1.1} \selectfont
    \includegraphics[width=0.9\columnwidth]{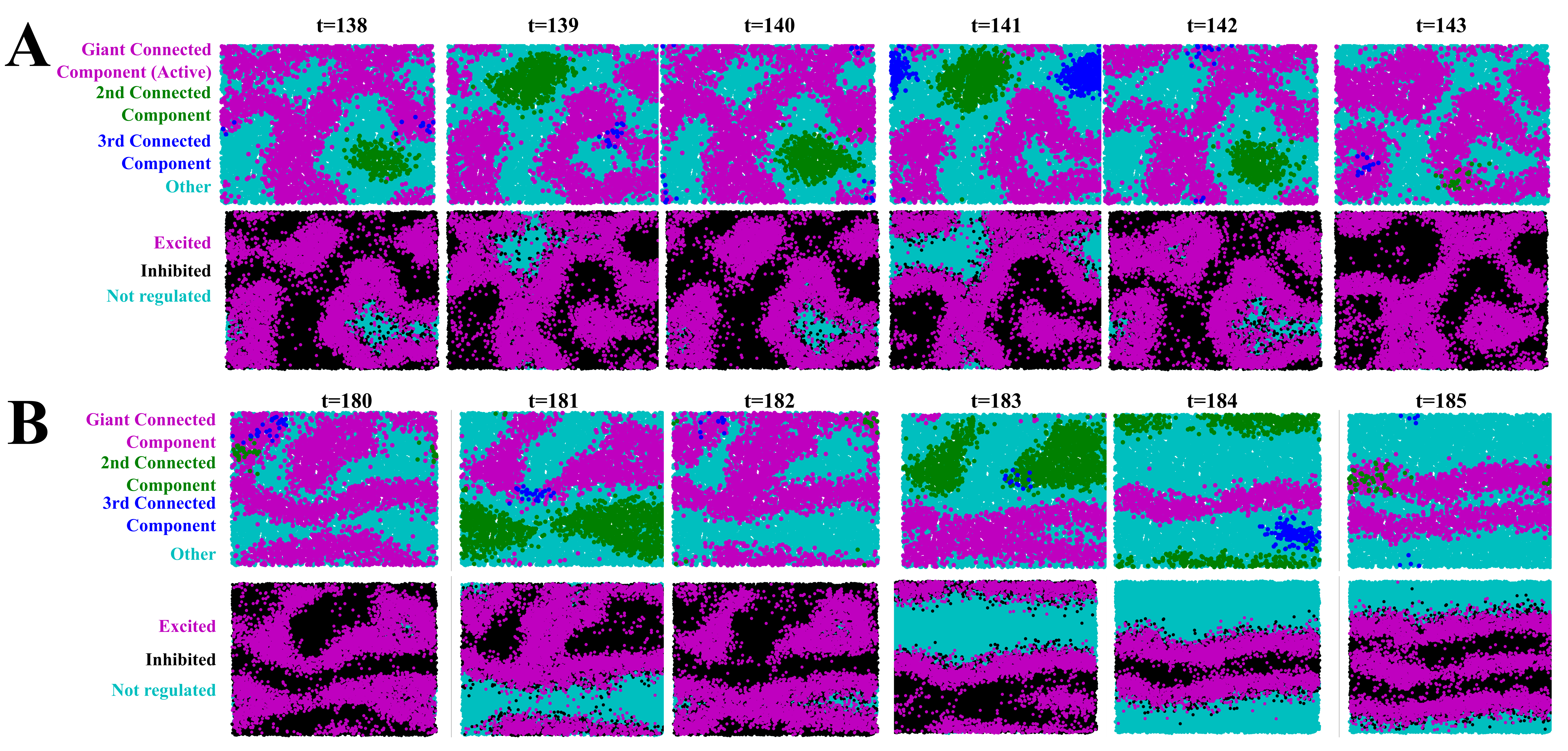}
    \caption{Mechanism underlying the emergence of spatial triadic percolation patterns. 
    Two exemplary snapshots into the evolution of the triadic percolation patterns (\textbf{A, B}) are shown. 
    The top panels (first and third row) show the states of the nodes, where active nodes (in the largest or giant connected component) are shown in pink as before, and we also show in green and blue the (inactive) nodes in the second and third largest connected components, for illustrative purposes. 
    The remaining inactive nodes are shown in light blue as before. 
    The bottom panels (second and fourth row) show the subsequent states of the structural links: up-regulated or excited (pink), down-regulated or inhibited (black), and not regulated (black). We do not show here random damage. 
    Excited links are defined as those that are exclusively up-regulated, whereas inhibited links comprise all down-regulated links, regardless of whether they are up-regulated or not, according to the activation rule. The spatial triadic network is formed by $N=10^3$ nodes, with parameter values $c=0.6$, $c^+=c^-=0.2$, $d_r=d_0=0.2$, $\rho=100$, $p=1.0$.
    }
    \label{fig:sup_fig_surface_tension}
\end{figure}

\begin{figure}[!tbh]
    \centering \linespread{1.1} \selectfont
    \includegraphics[width=0.9\columnwidth]{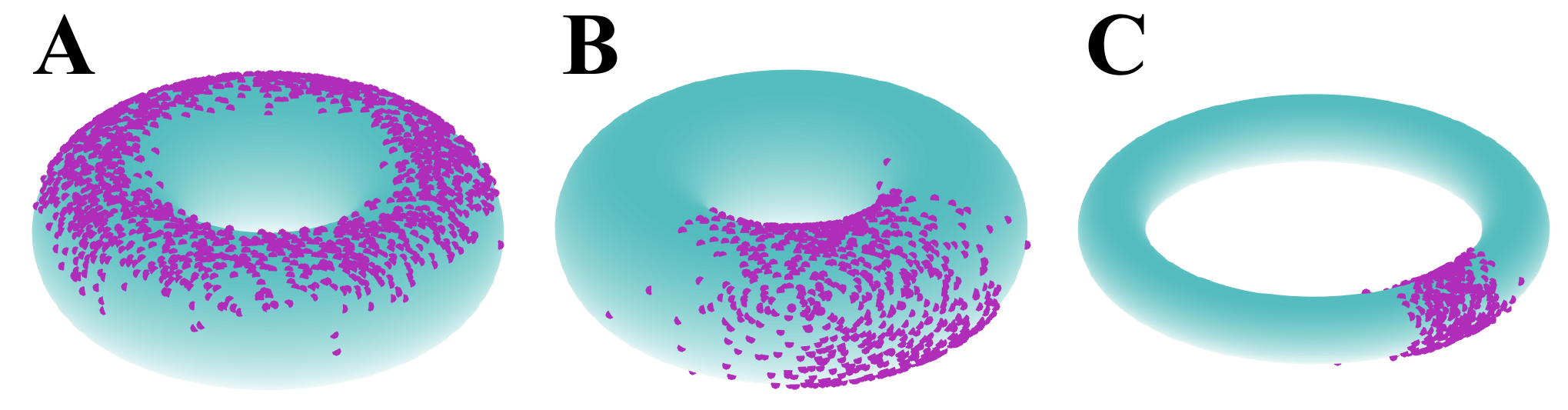}
    \caption{Emergent triadic percolation patterns for the square (\textbf{A}, \textbf{B}) and rectangular (\textbf{C})  lattice with periodic boundary conditions (tori). 
    The surface of the torus is shown with turquoise color, and active nodes are shown in pink. 
    In the square torus, the vertical and horizontal dimensions are symmetric, and in the deterministic regime $p=1$, both horizontal (A) and vertical (B) stripes emerge. 
    In the rectangular torus this symmetry is broken, and in the deterministic regime $p=1$  stripes emerge only in the shorter direction (e.g. vertical in this exemplary case). 
    The parameters for generating the spatial  networks with triadic interactions  for panel \textbf{A} and \textbf{B} are set as $N=10^4$, $c^{+} = c^{-} = 0.2$, $d_r = d_0 = 0.25$, $c = 0.6$. The size of the square $L=10$. 
    The parameters for panel \textbf{C} are  $N=10^4$, $c^{+}=c^{-}=0.2$, $d_r = d_0 = 0.25$, $c = 0.4$ and the rectangle has size $L_x = 22.36$, 
  $L_y=4.47$.}
    \label{fig:sup_torus}
\end{figure}

\subsection{Triadic percolation patterns: self-inhibition and border minimization}\label{sec:SI_spat}
In strongly spatial networks (small $d_0$ and $d_r$), the emergence of spatial triadic percolation patterns can be understood as a surface minimization effect caused by the underlying local positive and negative regulation mechanisms. 
Consider the pattern shown in the first panel of Supplementary Fig. $\ref{fig:sup_fig_surface_tension}$A (top image), corresponding to a strongly spatial network. Active nodes are shown in pink.
In the next step, active nodes will effect their regulation, resulting in structural links being either excited (positive regulation and no negative regulation), inhibited (negative regulation) or not regulated (no active regulatory links). The result of the regulation is shown in the bottom image of the panel, where pink, black and cyan points respectively show excited, inhibited and not-regulated links. 
The number of active positive and negative regulations of each link depends on its distance to the active regions. 
Thus, structural links within the active manifold typically have at least one active negative regulation and are inhibited (black points in the bottom panel of Supplementary Fig. $\ref{fig:sup_fig_surface_tension}$A). 
On the other hand, links far from the active regions have no active regulations and thus are not regulated (cyan points). 
Only at the border between the active and inactive node regions there can be sparse regulation, resulting in a set of links with only positive regulation, shown by the pink points (and another with negative regulation).
Consequently, the subsequent pattern (step $t+1$) roughly corresponds with the border of the current pattern (step $t$).
In this manner, the regulation acts as an effective surface tension, where the tension from inside of the active manifold is caused by negative regulation resulting in self-inhibition, and the tension from outside by the lack of positive regulation. 

\begin{figure}[!tbh]
    \centering \linespread{1.1} \selectfont
    \includegraphics[width=0.6\columnwidth]{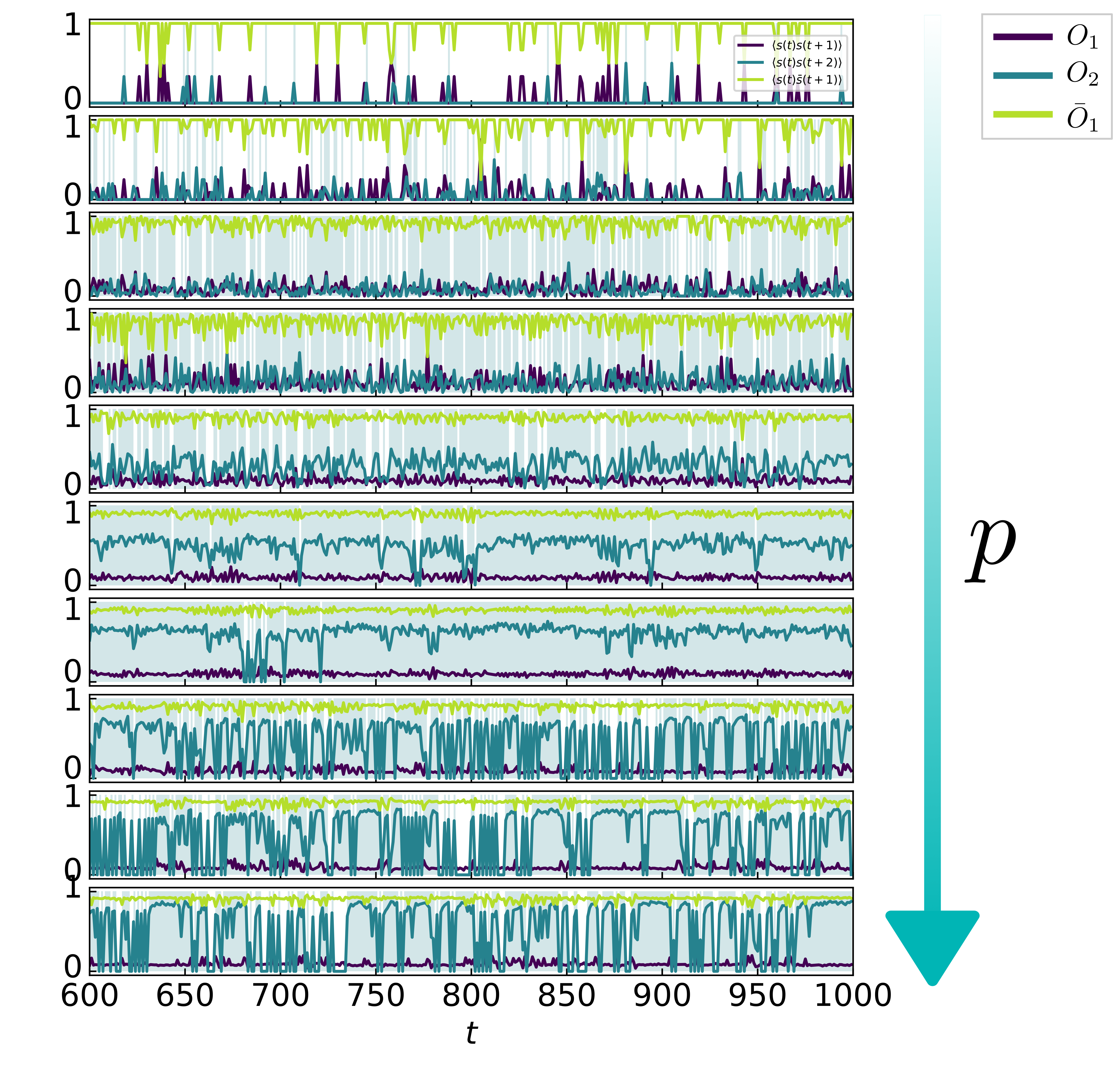}
    \caption{Temporal description: Overlap of the triadic percolation patterns.
    To illustrate the surface tension mechanism, we show the overlap parameters $O_1(t)$ (dark blue), $O_2(t)$ (turquoise) and $\bar{O}_1(t)$ (light green), for different values of $p=0.1,\ 0.2,\ ...,\ 1.0$, from top to bottom.
    Shaded areas indicate ST-blinking, i.e. $O_2(t)>\alpha$. These results are for the same network and realizations of the triadic percolation dynamics as those in Fig. $\ref{fig:figure6}$ of the main text, namely $N=10^4$, $c^+=c^-=0.2$, $d_r=d_0=0.25$, $c=0.4$, $\rho=100$.
    }
    \label{fig:sup_fig_overlap}
\end{figure}

Surface minimization explains the different observed behaviors. 
To quantify the temporal aspect, we have defined the overlap between patterns at different time-steps, namely
\bea
O_\tau(t) = \frac{\sum_i s_i(t) s_i(t+\tau)}{\sum_i s_i(t)},
\eea
and between the active and inactive patterns at different time-steps, i.e.
\bea
\bar{O}_\tau(t) = \frac{\sum_i s_i(t) \left(1-s_i(t+\tau)\right)}{\sum_i s_i(t)}.
\eea
Supplementary Fig. $\ref{fig:sup_fig_overlap}$ shows $O_1$, $O_2$ and $\bar{O}_1$ for $p=0.1,\ ...,\ 1.0$ for the same simulations are reported in Fig. $\ref{fig:figure6}$ of the main text.
For each pattern class, we observe the following dynamics:
\begin{enumerate}
    \item \textbf{Octopus patterns}. The active manifold and its neighbourhood conform the entire network, resulting in \emph{ST-blinking}.   
     \begin{enumerate} 
        \item The overlap between consecutive active states $O_1$ is close to $0$ (dark blue line in Supplementary Fig. $\ref{fig:sup_fig_overlap}$ for e.g. $p=0.7$), whereas $O_2$ is almost one (turquoise line) indicating an approximate period $2$ oscillation.
        \item $\bar{O}_1$ is large but smaller than $1$ ($0.5<\bar{O}_1<1.0$), indicating that pattern $t+1$ is typically the opposite of pattern $t$, but fluctuations occur.
        \item If pattern $t$ presents large loops, pattern $t+1$ can present two spatially separated active regions. 
        \item The patterns can become unstable due to quenched disorder or random damage (for $p<1$), causing either continuous deformations (Supplementary Fig. $\ref{fig:sup_fig_surface_tension}$A) or the break-down of part of the pattern (Supplementary Fig. $\ref{fig:sup_fig_surface_tension}$B).
    \end{enumerate}
    \item \textbf{Small clusters} \emph{diffuse} to nearby positions, and ST-blinking is not observed ($O_2\approx 0$). 
    \item \textbf{Stripes} emerge for large enough $p$ as they minimize the size of the border (Supplementary Fig. $\ref{fig:sup_fig_surface_tension}$B). 
    \begin{enumerate}
        \item Stripes can either \emph{ST-blink}  or \emph{diffuse}, see large fluctuations in $O_2$ for $p>0.7$ in Supplementary Fig. $\ref{fig:sup_fig_overlap}$. ST-blinking is characterized by large $O_2$, and diffusion by small $O_2$.  
        Note that the typical width of stripes is only one order of magnitude smaller than the size of the system (approx. $1/L$), and consequently effective diffusion is only possible on small time-scales before the stripe returns to the original region.     
        \item Multiple stripes can emerge from single stripes if the two neighbourhoods are connected by active structural links (Supplementary Fig. $\ref{fig:sup_fig_surface_tension}$B).  They require the activation of some longer distance structural links, and therefore are less stable and occur less often than single stripes for typical parameter values. Their emergence is most likely in the case of no random damage ($p=1$).       
    \end{enumerate}
\end{enumerate}

\subsection{Entropy rate}
 
Given a source producing an output that can be described through an ensemble of categorical time-series $x$, in order to calculate the entropy rate of $S$, $x$ is divided in non-overlapping motifs or words of length $\hat{L}$. 
Let $\tilde{p}_i$ be the normalized count of the $i$th word in the ensemble of words of length $\hat{L}$ in the time-series. 
Then, the estimate of the entropy rate (in bits per second or steps) is given by
\bea
H(l) = -\frac{1}{\hat{L}}\sum_i \tilde{p}_i \log_2 \tilde{p}_i.
\eea
The true entropy rate $\hat{H}$ of the source is reached in the limit of infinitely long words, i.e.
\bea \label{eq:erate_entropy}
\hat{H} = \lim_{\hat{L}\to \infty} H(\hat{L}).
\eea
In practice, the entropy rate of the source is estimated by averaging over the entropy rate of an ensemble of individual time-series $x$, $\hat{H}_x$:
\bea 
\hat{H} = \langle \hat{H}_x \rangle.
\eea

The entropy is a property of sources, and its estimation requires extensive sampling and long time-series that are not always available. Ref. \cite{amigo2004estimating} proposed an alternative estimator based on the Lempel-Ziv complexity $LZ76$ \cite{lempel1976complexity}. 
The $LZ76$ is defined recursively. 
Given the sequence $x_1^n:=x_1x_2,...,x_n$ of length $n$ ($1\leq i \leq n$), a block of length $n'$, $B_{n'}$, is a segment of length $n'$ of $x_1^n$, i.e. $x_{i+1}^{i+n'}:=x_{i+1}x_{i+2}... x_{i+n'}$. 
We set $B_1=x_1^1=x_1$. Suppose that 
\bea
x_1^{n_k} = B_1 B_2,...,B_k.
\eea
Then, we define 
\bea
B_{k+1}:= x_{n_k+1}^{n_{k+1}} \left( n_k+1 \leq n_{k+1} \leq n \right),
\eea
to be the block of minimal length that does not occur in the sequence $x_1^{n_{k+1}-1}$. 
By iterating this procedure, we decompose $x_1^n$ in minimal blocks, 
\bea
x_1^n = B_1 B_2,...,B_p.
\eea
in which all blocks are unique except for (potentially) the last block $B_p$. The $LZ76$ complexity of $x_1^n$ is defined as the number of blocks in the decomposition (which is unique):
\bea
C(x_1^n) := p.
\eea
The rate of generation of new patterns along $x_1^n$ is measured by the normalized $LZ76$ complexity
\bea
c(x_1^n) = \frac{C(x_1^n)}{n/\log_2 n} = \frac{p}{n}\log_2 n.
\eea
Sequences that are not complex have a very small normalized $LZ76$ complexity, and random sequences have maximal $LZ76$ complexity. 
If a source is stationary, then \cite{ziv1978coding, amigo2004estimating}
\bea \label{eq:lim_H_c_stat}
\limsup_{n\to \infty} c(x_1^n) \leq H,
\eea
on average and, if the source is ergodic, then
\bea \label{eq:erate_LZ76}
\limsup_{n\to \infty} c(x_1^n) = H,
\eea
almost surely.
These equations provide estimates from below to the entropy of a source through the $LZ76$ complexity of the generated time-series.
Note that whereas the limit to infinity in Eq.  ($\ref{eq:erate_entropy}$) is on the word length, which inevitably leads to under-sampling of long words and requires extrapolating the results of short words, the limit in Eq. ($\ref{eq:erate_LZ76}$) is on the length of the time-series and, if the converge of Eq. ($\ref{eq:erate_LZ76}$), a reliable estimate can be achieved even in short time-series.

In Supplementary Fig. $\ref{fig:sup_fig_entropy_rate}$ we show the two procedures described above to estimate the entropy rate, for a set of values of $p$. Namely, in panel A we show the entropy rate as derived from the normalized $LZ76$ complexity $c(x_1^n)$ when considering an increasing length $n$ of the time-series. As it can be shown, convergence is reached for most values of $p$. The entropy rate is then measured in this case as $c(x_1^{n=T})$, where $T$ is the length of the time-series.
In panel B we show the entropy rate $H(\hat{L})$ of words of increasing length $\hat{L}$ as estimated via Eq. ($\ref{eq:erate_entropy}$). 
The true entropy rate is measured in this case by performing a linear fit to the last two points of the plot (corresponding to $\hat{L}=1,\ 2$) and extrapolating the trend to $1/\hat{L}=0$. 
The results of both estimates are shown in panel C for the same data shown in panels A and B, and in panel D for an ensemble average over $10$ iterations of the triadic percolation system and dynamics. As it can be seen, both methods lead to the same results for the cluster-dominated regime (up to the first maximum), but there is a significant deviation between them for the octopus-stripe co-existence region (region around the second maximum). According to Eqs. ($\ref{eq:lim_H_c_stat}$) and ($\ref{eq:erate_LZ76}$), this result points towards the non-ergodicity of the pattern time-series, although more detailed analyses would be required to validate this finding. 

\begin{figure}[!tbh]
    \centering \linespread{1.1} \selectfont
    \includegraphics[width=0.8\columnwidth]{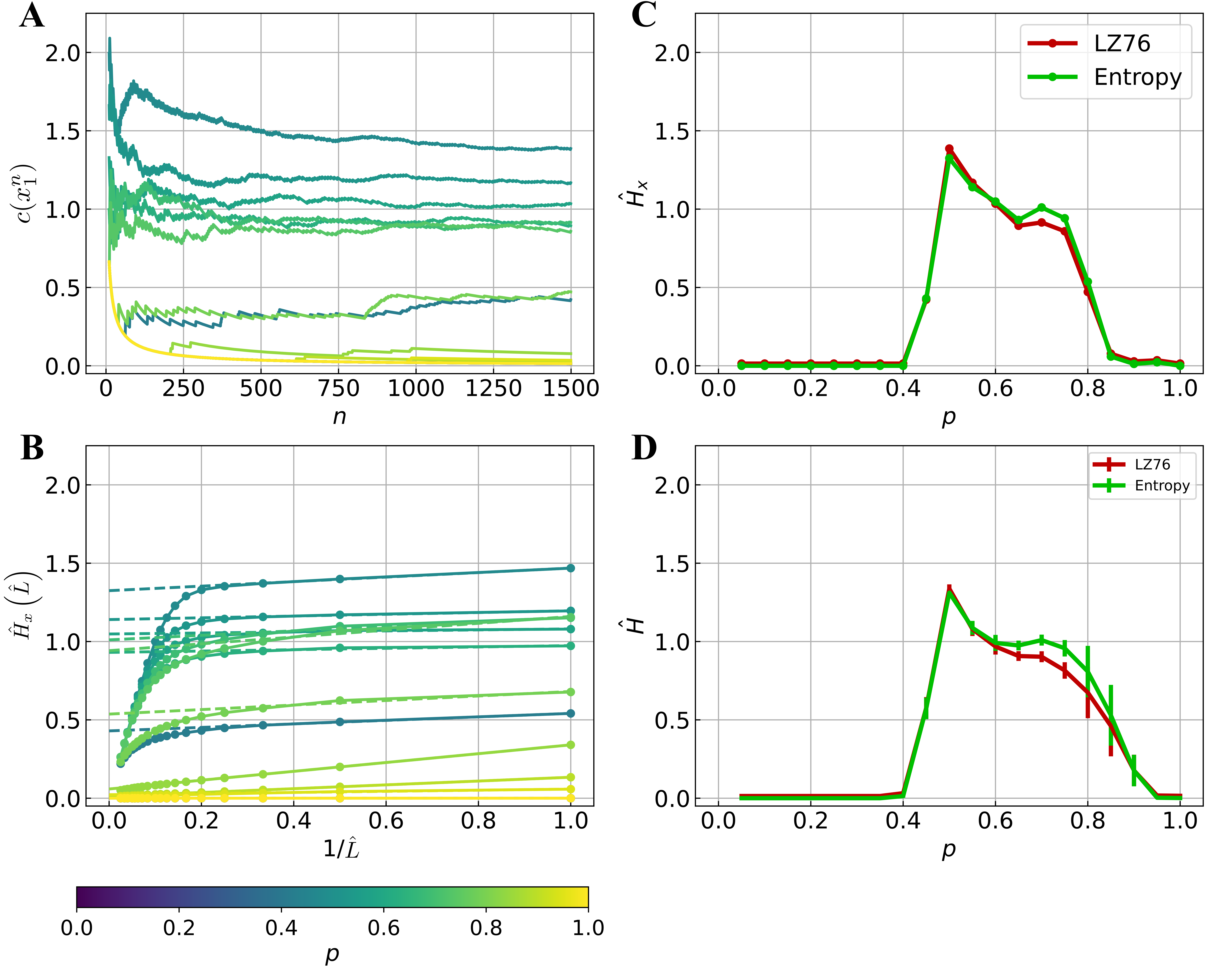}
    \caption{Entropy rate estimation. Two procedures were considered to estimate the entropy rate, following Ref. \cite{amigo2004estimating}. 
    \textbf{A} Normalized Lempel-Ziv $76$ complexity $LZ76$ \cite{lempel1976complexity, kaspar1987easily}, $c(x_1^n)$, of the pattern time-series $x$ for increasing length $n$. Each line corresponds to a single time-series generated over the same system with different probabilities of random damage $p$ as indicated by the colorbar. The limit $n\to \infty$ gives the entropy rate, which is estimated here as $c(x_1^{n=T})$ \cite{amigo2004estimating}, where $T=1500$ is the integration time. 
    \textbf{B} Entropy rate  $\hat{H}_x^{\hat{L}}$ of the pattern time-series $x$ when considering motifs or words of increasing length $\hat{L}$. To estimate the entropy rate as $\hat{L}\to \infty$, the linear trend on the right-hand side of the plot is extrapolated to $1/\hat{L} \to 0$ \cite{amigo2004estimating}. The data in panel B was generated over the same set of time-series as in panel A.
    \textbf{C} The two estimates of the entropy rate $c(x_1^{n=T})$ and $\hat{H}^{\hat{L}\to \infty}_x$ are shown as a function of the random damage probability $p$, as indicated by the legend, for the data shown in panels A and B. \textbf{D} Ensemble-average of the two estimates of $\hat{H}$, averaged over $10$ realizations of the triadic percolation system and dynamics for each value of $p$. The errorbars indicate the standard deviation. In all panels, the spatial network with triadic interactions is formed by $N=10^4$ nodes and generated with parameter values  $c^+=c^-=0.2$, $d_r=d_0=0.25$, $c=0.4$, and we considered a simulation length of $2000$ steps. The first $500$ steps were discarded as transient.}
    \label{fig:sup_fig_entropy_rate}
\end{figure}

\subsection{Rectangular torus}
We consider here the effect of breaking the horizontal-vertical symmetry in the emerging  spatial triadic percolation patterns.
In order to do so, we simulate the triadic percolation model on a rectangle (width five times its height) with periodic boundary conditions (rectangular torus), whilst all other network and model parameters (in particular the number of nodes, density and scaling of connections with the distance) are kept the same. 
As shown in Supplementary Fig. $\ref{fig:sup_torus}$, stripes always emerge along the shorter dimension in the rectangular torus (see panel C). 
This is in agreement with the proposed mechanism for the emergence of macroscopic activity patterns, based on the minimization of surface tension. 
Moreover, we have found that stripes emerge for higher noise values (smaller $p$) in the case of the rectangular torus, indicating that shorter stripes are more stable.

\subsection{Barycenter dynamics}

For stripe patterns the barycenter in the perpendicular dimension is well-defined. To determine whether stripes predominantly ST-blink or diffuse, we have analyzed the evolution of the barycenter. As a reference, we include also the study of the barycenter for clusters, which can also be defined, and in which case the overlap analysis indicates a diffusive behavior. We consider the case of horizontal stripes and thus focus on the vertical component of the barycenter location.

In Supplementary Fig. $\ref{fig:sup_fig_barycenter}$A and D we show a recurrence plot \cite{eckmann1995recurrence,marwan2007recurrence} of the barycenter location along the vertical axis, $y_B(t)$, respectively for small clusters and stripes. We observe that for small clusters the barycenter moves in time through the network, and all locations are reached with homogeneous probability. 
All points in the recurrence plot fall along the diagonal, indicating a short displacement in each step, due to the small size of the patterns. 
The situation for stripes is vastly different, and only some locations are visited by the pattern. 
Moreover, there is a strong preference for a handful of locations. 
At the microscopic scale, however, deviations of the barycenter position within each macroscopic preferred location are observed over time, indicating that the actual stripe changes over time (in agreement with Fig. $\ref{fig:figure4_timeseries_RHC_TDA}$D). 
Points in the recurrence plot fall further from the diagonal in the case of stripes, as these are wider than clusters and the patterns move further on each step. 
Moreover, the displacements of stripes typically take place in discrete amounts, i.e. stripes typically move the width of one,  and occasionally two (in the case of multiple stripes), stripes.

To discriminate whether there is asymmetry in the movement of the patterns,
in Supplementary Fig. $\ref{fig:sup_fig_barycenter}$ we display the asymmetry in the movement for each location $y$, derived from an analysis of the barycenter recurrence plot \cite{eckmann1995recurrence, marwan2007recurrence}. 
In order to do so, we have measured the number of times that the pattern moves up ($N_U$) and down ($N_D$) from each location, and report this by the blue and red curves, respectively. 
Then, the asymmetry at each location $A(y)$ is measured as the difference between $N_U$ and $N_D$ (black curve in the bottom panels). 
To measure $A(y)$ we made use of a coarse-grained grid of size $30\times30$, equal to the one used for the spatial characterization of the patterns in the main text.
No significant asymmetry is found for clusters, whereas for stripes there are two distinct locations, with positive and negative asymmetry respectively, such that the stripe ST-blinks between them (i.e. there is a large probability of moving upwards from the lower location, and conversely of moving downwards from the upper position). 
This asymmetry arises due to quenched disorder in the structural and regulatory networks, and leads to ST-blinking being the preferred mechanism for $p=1$. We note that this dynamics cannot be observed clearly through $R(t)$, or through $H(t)$ or $C(t)$, as stripes are typically of the same size and shape. 

Finally, we have measured the 
return probability $P_0(\tau)$ of the barycenter to a given location after $\tau$ steps (see Supplementary Fig. $\ref{fig:sup_fig_barycenter}$C, F respectively for clusters and stripes).
A return event after $\tau$ steps occurs if $x_B(t)$ and $x_B(t+\tau)$ belong to the same cell of the coarse-grained $30\times 30$ grid.
$P_0(\tau)$ confirms that clusters show a power-law decay of $P_0(\tau)$ on short to intermediate time-scales, until a plateau is reached. 
On the contrary, the decay for stripes is slower and a plateau is reached earlier. We also notice a significant difference for odd and even times for stripes due to the self-inhibition effect. 

In summary, diffusion only emerges at short time-scales, due to quenched disorder caused by local inhomogeneities. 
In general, the probabilities of moving upwards and downwards (or right and left in the case of vertical stripes) are not equal and depend on the position, effectively trapping the stripe into preferred locations with only short visits to other network regions, resulting in an inhomogeneous spatial distribution of activity over time.

\begin{figure}[!tbh]
    \centering \linespread{1.1} \selectfont
    \includegraphics[width=0.8\columnwidth]{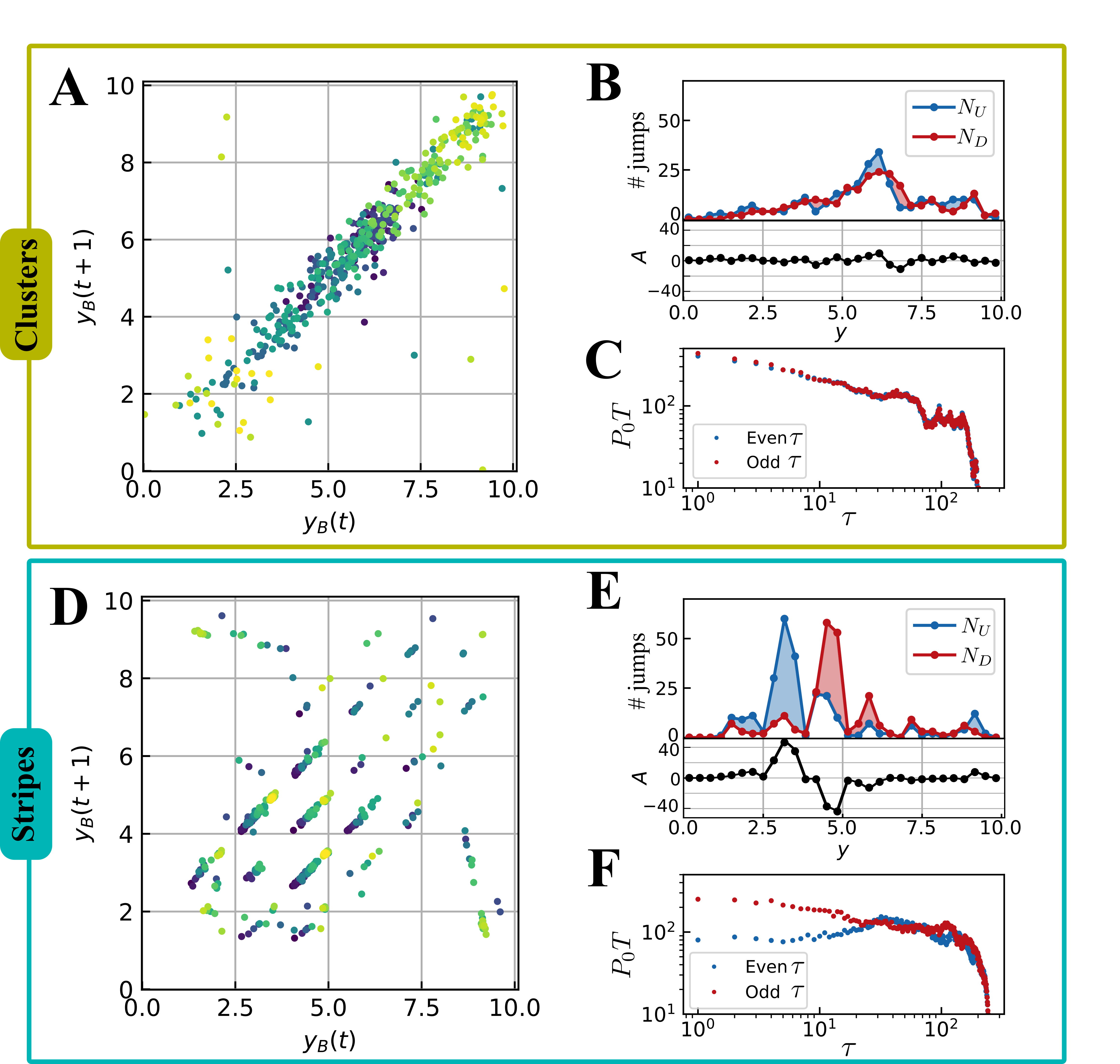}
    \caption{Barycenter dynamics of the spatial triadic percolation patterns along the vertical axis corresponding to small clusters ($p=0.3$, panels A-D) and horizontal stripes ($p=1.0$, panels E-H).  
    \textbf{A}, \textbf{D} Recurrence plot of the barycenter location $y_B$ with lag $\tau=1$. Each point corresponds to a time-step, with the time indicated by the color scale (from dark blue for $t=600$ to yellow for $t=1000$).
    \textbf{B}, \textbf{E} Asymmetry in the movement of the barycenter. The top plot shows the number of times the barycenter moves up ($N_U$, blue points) and down ($N_D$, red points) from each position. 
    The bottom plot shows the asymmetry at each position, $A(y)$, defined as $A(y) = N_U(y)-N_D(y)$. 
    \textbf{C}, \textbf{F} Number of return events $P_0(\tau)T$ (i.e. the return probability $P_0(\tau)$ times the total number of steps considered $T$). 
    Odd and even times have been separated as they follow different scaling for stripes due to the self-inhibition effect. These results are for the same network and realizations of the triadic percolation dynamics as those in Fig. $\ref{fig:figure6}$ of the main text, namely $N=10^4$, $c^+=c^-=0.2$, $d_r=d_0=0.25$, $c=0.4$, $\rho=100$, $T=400$.
    }
    \label{fig:sup_fig_barycenter}
\end{figure}


\end{document}